\documentclass{naturesubmissionstyle}

\makeatletter
\renewcommand\@biblabel[1]{#1.}
\makeatother

\usepackage{multirow}
\usepackage{graphicx}
\usepackage[super,comma]{natbib}
\usepackage[resetlabels]{multibib}
\newcommand\aap{A\&A}                
\newcommand\apj{ApJ}                 
\newcommand\apjl{ApJ}                
\newcommand\araa{ARA\&A}             
\newcommand\mnras{MNRAS}             
\newcommand\nar{New~Astron.~Rev.}    
\newcommand\nat{Nature}              
\newcommand\prd{Phys. Rev.~D}        
\newcommand\prl{Phys. Rev.~Lett.}    
\newcommand\pasp{PASP}               

\usepackage{amssymb}
\usepackage[hypertexnames=false,colorlinks]{hyperref}

\usepackage{amsmath}
\usepackage{soul}

\usepackage{url}
\usepackage{mathtools}
\usepackage{physics}
\usepackage[dvipsnames]{xcolor}

\usepackage{soul}
\usepackage{lineno}

\title{Exceptionally bright optical emission from a rare and distant $\rm \gamma-$ray burst}
\author{Gor Oganesyan$^{1,2}$, Sergey Karpov$^{4,5,6}$, Martin Jel\'{\i}nek$^{7}$, Gregory Beskin$^{5,6}$, Samuele Ronchini$^{1,2}$, Biswajit Banerjee$^{1,2}$,
Marica Branchesi$^{1,2,3}$,
Jan \v{S}trobl$^7$, Cyril Pol\'{a}\v{s}ek$^7$, Ren\'{e} Hudec$^7$,
Eugeny Ivanov$^{8}$, Elena Katkova$^{8}$, 
Alexey Perkov$^{8}$, Anton Biryukov$^{9,6}$, Nadezhda Lyapshina$^5$
Vyacheslav Sasyuk$^6$, Martin Ma\v{s}ek$^{4}$, Petr Jane\v{c}ek$^{4}$, Jan Ebr$^{4}$, 
Jakub Jury\v{s}ek$^{4}$, Ronan Cunniffe$^{4}$, Michael Prouza$^{4}$}

\date{}

\begin{document}

\maketitle
\let\thefootnote\relax\footnote{
\begin{affiliations}
\item Gran Sasso Science Institute, Viale F. Crispi 7, I-67100, L’Aquila (AQ), Italy 
\item INFN - Laboratori Nazionali del Gran Sasso, I-67100, L’Aquila (AQ), Italy
\item INAF - Osservatorio Astronomico d’Abruzzo, Via M. Maggini snc, I-64100 Teramo, Italy
\item CEICO, Institute of Physics, Czech Academy of Sciences, Prague, Czech Republic. 
\item Special Astrophysical Observatory, Russian Academy of Sciences, Nizhniy Arkhyz, Russia
\item Kazan Federal University, Kazan, Russia
\item Astronomical Institute, Czech Academy of Sciences (ASU CAS), Fri\v{c}ova 298, Ond\v{r}ejov, Czech Republic
\item OJS RPC PSI, Nizhniy Arkhyz, Russia
\item Sternberg Astronomical Institute, Moscow State University, Russia
\end{affiliations}
}

\baselineskip24pt

{\noindent \bf Long $\rm \gamma$-ray bursts (GRBs) are produced by the dissipation of ultra-relativistic jets launched by newly-born black holes after the collapse of massive stars. Right after the luminous and highly variable $\gamma$-ray emission, the multi-wavelength afterglow is released by the external dissipation of the jet in circumburst medium. We report the discovery of a very bright ($\rm \sim 10$ mag) optical emission 
$\rm \sim 28$ s after the
explosion of the extremely luminous and energetic GRB 210619B located at redshift 1.937.  
Early multi-filter observations allowed us to witness the end of the shock wave propagation into the GRB ejecta. We 
observed the spectral transition from a bright reverse to the forward shock emission, 
demonstrating that
the early and late GRB multi-wavelength emission is originated from a very narrow jet propagating into an unusually rarefied interstellar medium. We also find evidence of an additional component of radiation, coming from the jet wings which is able explain the uncorrelated optical/X-ray emission.  

}

\newpage 

\begin{figure}
    \centering
    \includegraphics[width=0.80\textwidth]{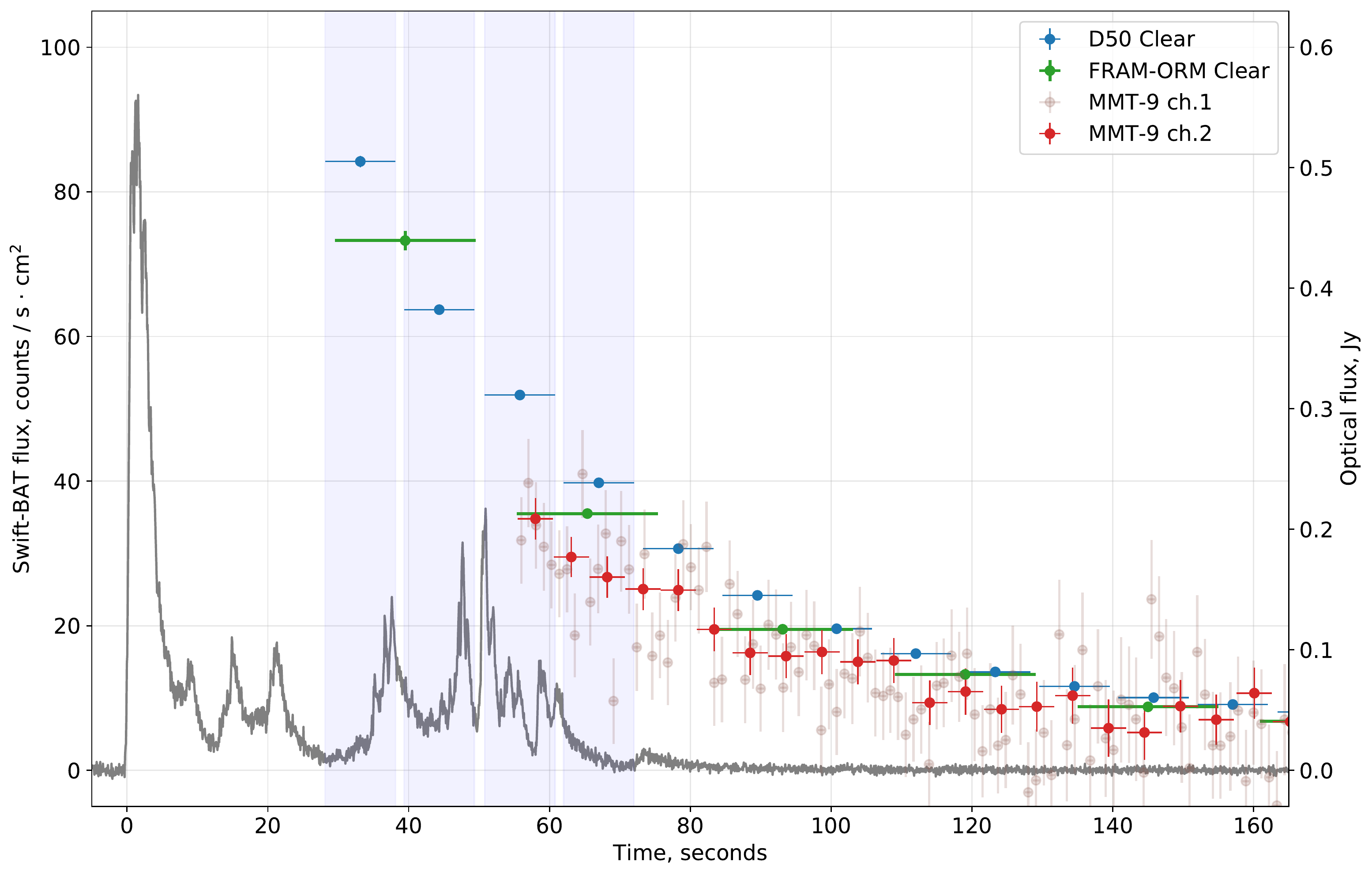}
    \includegraphics[width=0.85\textwidth]{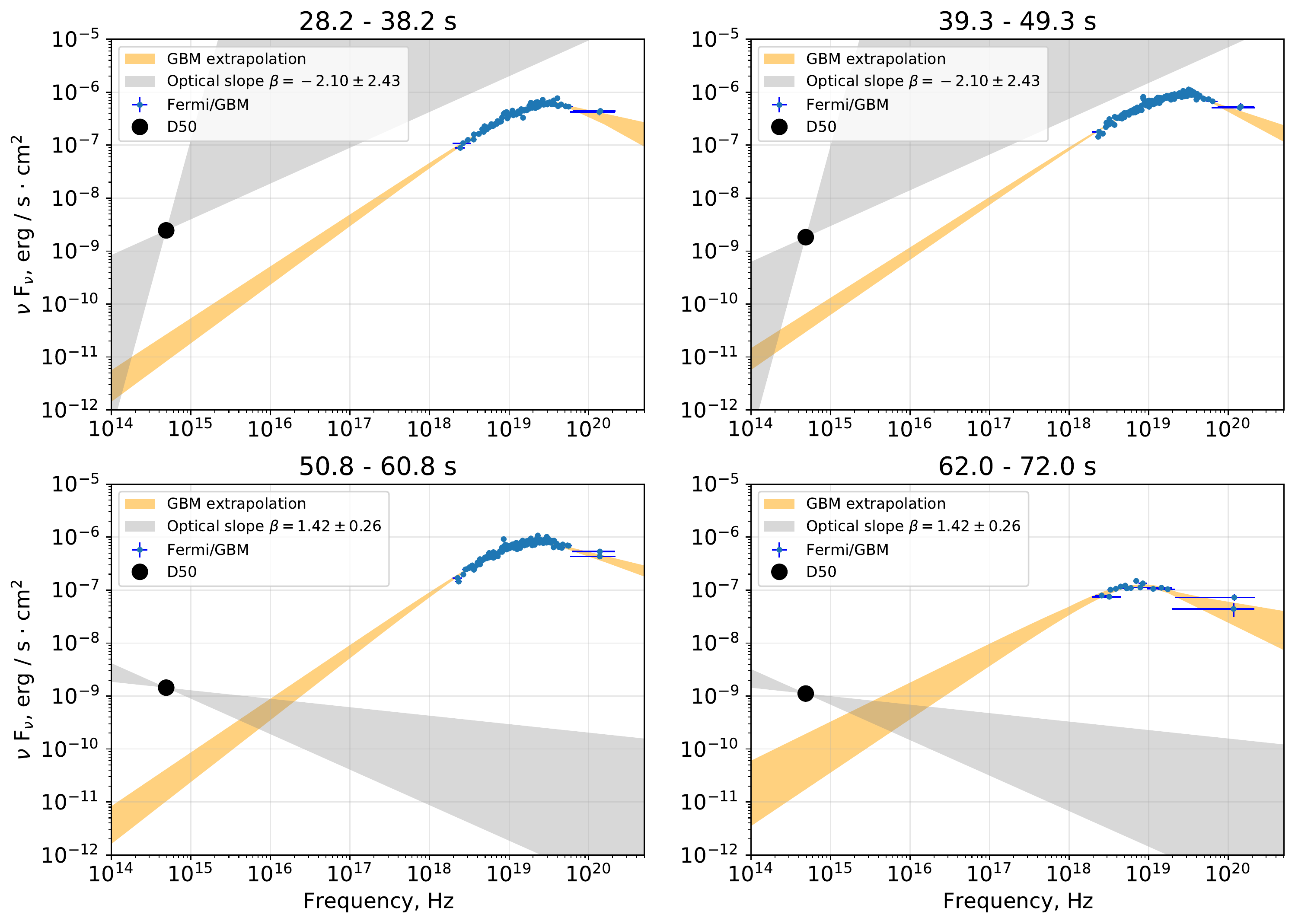}
    \caption{Upper panel -- early optical lightcurves from D50, FRAM-ORM and Mini-MegaTORTORA (channels with 1 s and 5 s temporal resolution) telescopes along with Swift-BAT count rate. Optical fluxes are corrected for extinction. Blue vertical bars denote the intervals for whose joint gamma-optical spectra are shown on lower panels. Lower panels -- Fermi GBM spectrum and its extrapolation towards optical range at different time intervals along with optical flux measured by D50. 
    We also show the slope of the optical spectrum. The slope is estimated using data from several telescopes observing simultaneously in slightly different wavebands.
    }
    \label{fig:gamma_opt}
\end{figure}

On June 19, 2021, at 23:59:25 Universal Time (UT) the Swift Burst Alert Telescope triggered on GRB~210619B\cite{gcn_swift_bat} 
and promptly distributed its coordinates. GRB~210619B was also detected by the Gravitational wave high energy Electromagnetic Counterpart All sky Monitor\cite{gcn_gecam}, the Konus/WIND Experiment\cite{gcn_konus}, and by the Fermi Gamma-Ray Burst Monitor\cite{gcn_fermi}. This allowed ground-based robotic telescopes to rapidly repoint towards it, starting unfiltered optical observations while gamma-ray activity was still ongoing. D50\cite{d50,gcn_d50} and FRAM-ORM\cite{fram_la_palma} telescopes were able to capture the earliest optical emission. The data acquired 
since $T_0+28$ s display an overall smooth optical light curve
which faded from magnitude $r=10.6$ to $r=17$ about 3000 seconds later as shown in Fig.~\ref{fig:gamma_opt}. Prompt observations by Mini-MegaTORTORA (MMT-9) camera\cite{beskin_mmt_2017}
started at $T_0+55.5$ s and 
provided 
high temporal resolution (1 s exposures) data and images of the transient in photometric Johnson B and V filters. 

\begin{figure}
    \centering
    \includegraphics[width=1.0\textwidth]{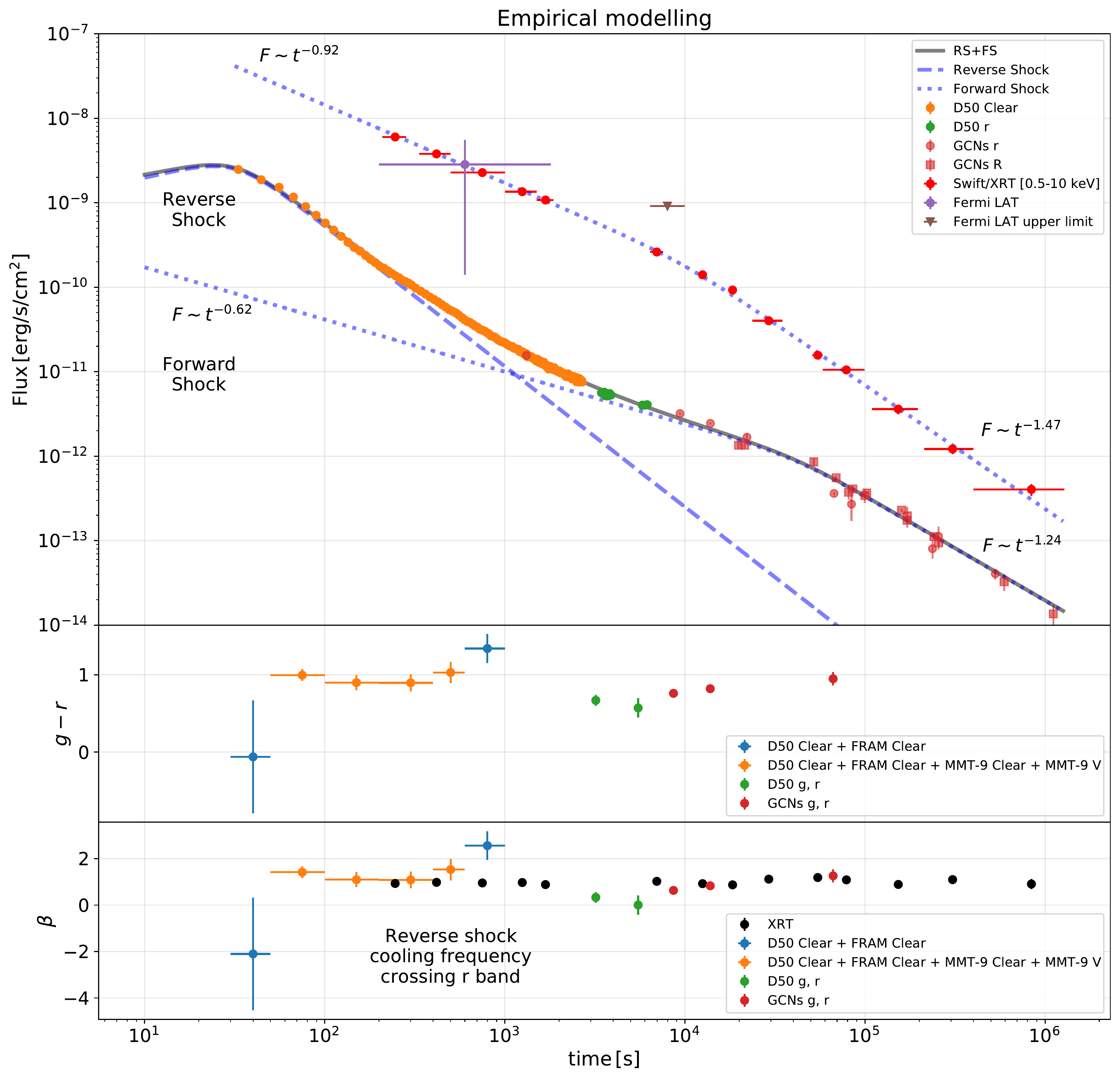}
    \caption{Upper panel -- overall optical (D50) and X-ray (Swift XRT) light curves and their empirical decomposition into earlier reverse shock and later forward shock components. Middle panel -- optical $g-r$ color estimated from comparison of fluxes from simultaneous unfiltered observations by several telescopes (D50, FRAM-ORM, and Mini-MegaTORTORA, earlier points) and from direct measurements in photometric filters (D50 and data from GCN circulars, later points). The measurements are not corrected for Galactic and extragalactic extinction. Lower panel -- corresponding $F=\nu^{-\beta}$ spectral slopes after correcting for the Galactic extinction (with $E_{g-r}^G=0.17$) and an additional extragalactic intervening dust absorber at $z=1.095$ (with $E_{g-r}=0.40$).}
    \label{fig:lc}
\end{figure}

These observations allowed us to probe the color evolution (see Fig.~\ref{fig:lc}) of the GRB~210619B optical afterglow over the first hundreds of seconds since its onset.
Following observations with the Gran Telescopio CANARIAS (GTC) revealed\cite{gcn_redshift} the redshift of the burst, $z=1.937$ placing the event among the most luminous ($\rm \sim 2.5 \times 10^{54}$erg/s) and most energetic ($\rm \sim 3.4 \times 10^{54}$ erg) ever detected (see Methods). 

The prompt emission which lasts for $\rm \sim 80$ s is thought to originate from the internal dissipation of the jet via shocks\cite{Narayan1992,Rees1994} or magnetic re-connection\cite{Drenkhahn2002,Lyutikov2003,Zhang2011} in an optically thin region. The dominant contribution to the prompt emission is suggested to be the synchrotron radiation from non thermal population of electrons\cite{Rees1994,Sari1996}. 
To establish whether the observed early optical emission is originated from the prompt emission, we 
performed time-resolved spectral analysis of the prompt emission 
using the time-bins corresponding to the data-sets provided by D50 and MMT-9. To model the GRB spectra in the widest possible energy range, we select the data provided by the Fermi/GBM instrument (8 keV - 10 MeV). For all the time-resolved spectra of Fermi/GBM we find that the optical flux exceeds the low-energy extrapolation of the GRB spectra by 2-2.5 orders of magnitude (see Fig.~\ref{fig:gamma_opt}). This analysis excludes the possibility for the optical emission to arise from a single-component prompt emission spectrum\cite{Oganesyan2019}. We also exclude that the optical emission could have been produced by an additional low-energy component in the prompt phase for the following reasons. In a standard optically thin emission regions, the bright optical emission would originate from a synchrotron cooling of non-thermal electrons, while the usual keV-MeV emission -- as an Inverse Compton radiation from the same seed photons. In this model, the second order Inverse Compton component would be brighter than the keV-MeV emission by itself causing an energy budget problem\cite{Derishev2001,Piran2009}. Alternatively, the 30-100 s optical emission could be the delayed prompt emission from the first bright episode of the GRB at 0-20 s. This would require a mechanism that can store some part of the energy of the prompt emission and release it later, at larger distances. High energy neutrons, primarily present in the fireball\cite{Derishev1999,Fuller2000,Beloborodov2003} or produced on the prompt emission side, could in principle provide this mechanism by their later decay. However, the bulk Lorentz factor required in this scenario is rather too small $\rm \Gamma \sim 25 $\cite{Fan2009} (for a delay of 50 s) compared to the Lorentz factor distribution of long GRBs\cite{Ghirlanda2018}. 

\begin{figure}
    \centering
    \includegraphics[width=0.9\textwidth]{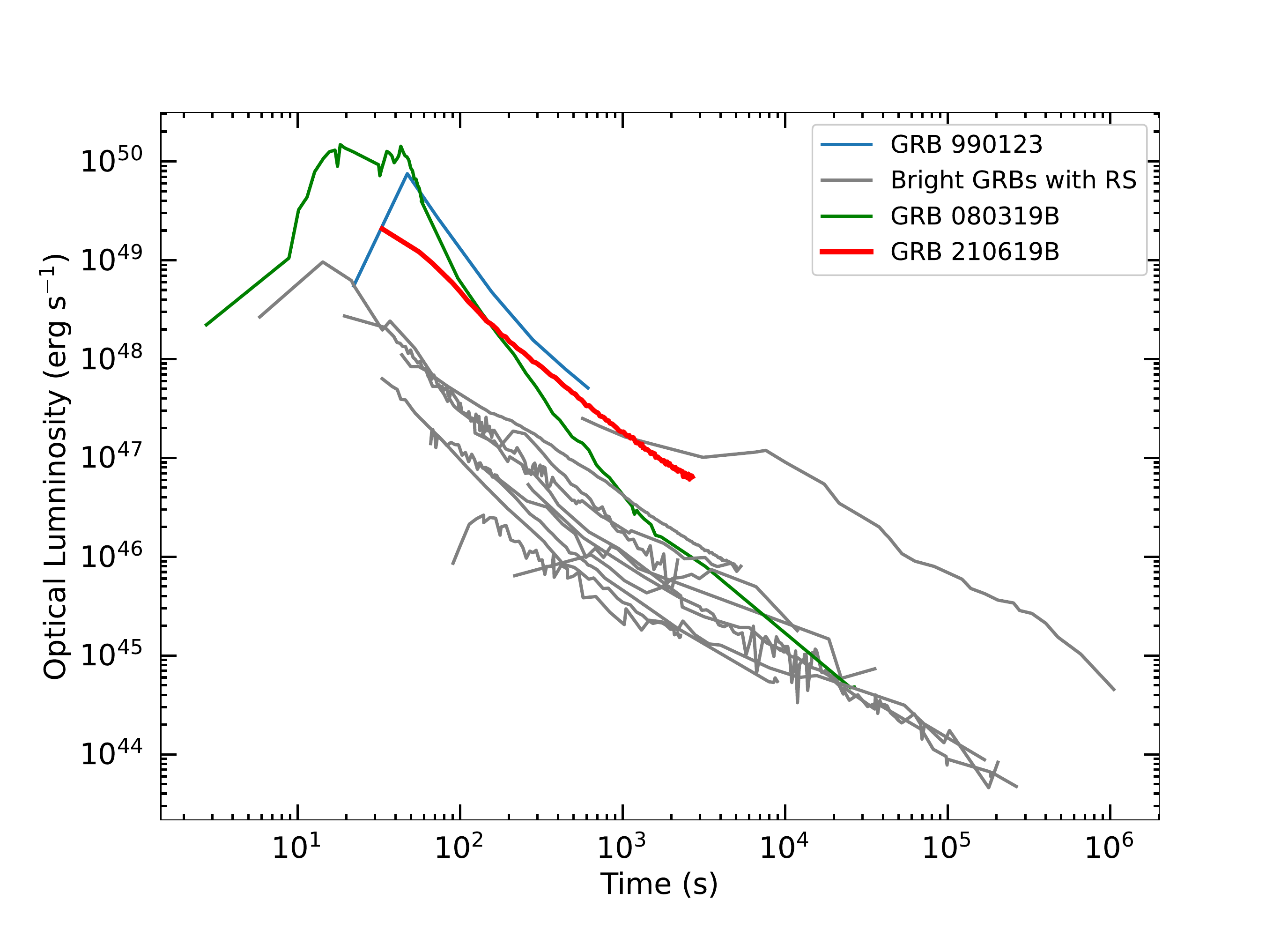}
    \caption{Comparison between optical light curves of GRB 210619B (red solid line) and other GRBs with bright reverse shock component (except GRB 080319B). Two bright GRBs, namely, GRB 990123\cite{Akerlof1999} and GRB 080319B\cite{Racusin2008} are shown with blue and green solid lines, respectively. For GRB 080319B, data before and after 60\,s are from the optical V-band and optical R-band, respectively. The grey solid lines represent the GRBs (GRB 021004\cite{021004}, GRB 021211\cite{021211},GRB 050525A \cite{050525A}, GRB 061126\cite{061126}, GRB 081007\cite{081007}, GRB 090102\cite{090102}, GRB 091024\cite{091024,gcn_10066,gcn_10073,gcn_10074}, GRB 130427A\cite{130427A}, GRB 140102A\cite{140102A}, GRB 140512A\cite{140512A}, and GRB 190114C\cite{190114C}) observed with (possible) RS component.}
    \label{fig:all_RS}
\end{figure}

A bright and primarily optical flash is expected from a reverse shock (RS, hereafter) propagating into the GRB ejecta\cite{Meszaros1997,Sari1999,Kobayashi2000}. The RS heats the particles of the ejecta and 
fades immediately after crossing them.
Due to the energy sharing among 
densely populated particles, the average energy gain of an electron is rather small compared to the forward shock scenario, making the peak of the RS emission to appear in the optical, infrared and radio bands. An order of $\sim 20$ optical counterparts of GRBs (see Fig.~\ref{fig:all_RS}) and few late-time radio transients were suggested to originate from RS \cite{Laskar2013,Laskar2019a,Laskar2019b}. 
Figure~\ref{fig:all_RS} shows the comparison among the early optical emission of GRB 210619B and all the bright optical 
emission interpreted as RS for other GRBs, along with the optical light curve of the highly variable naked-eye GRB 080319B\cite{Racusin2008,Beskin2010}. 
The RS candidate from GRB 210619B is the most luminous together with the historical GRB 990123\cite{Akerlof1999}. However, compared to GRB 990123 which is characterised by few optical exposures, the optical light curve of GRB 210619B is the first observed in great details allowing us to accurately model it by external shocks.

We first model the optical and X-ray light curves separately by series of power-laws (See Methods). The optical light curve is best represented by the RS component that peaks at $\rm \sim 30$ s with a following decay of $\rm t^{-1.6}$. Forward shock (FS, hereafter) is initially subdominant but it becomes later the only component in the optical band. The FS is best described by $\rm t^{-0.6}$ before the break time of $\rm \sim 4 \times 10^{4}$ s and $\rm t^{-1.2}$ later on.
The X-ray light curve is simply composed by two power-laws smoothly connected at $\rm \sim 10^{4}$ s with the shapes of $\rm t^{-0.9}$ and $\rm t^{-1.5}$ before and after. We preliminary interpret the optical/X-ray lightcurves as arising from the RS (only in the optical band) and the FS with an early jet-break at $\rm 10^{4}$ s. The temporal shapes and the optical/X-ray spectral indices ($\rm \beta_{opt} \approx \beta_{X} \approx 0.9$, where $\rm F_{\nu,opt,X} \propto \nu^{-\beta}$) suggest that at least before the jet break the observed cooling frequency of the FS is above the X-ray band (0.5-10 keV). This is additionally supported by the measure of the Fermi/LAT ($\rm >$100 MeV) spectral index of $\rm \beta_{LAT}=1.68 \pm 0.32$. However, a detailed investigation of the shapes of the light curves suggests that the usual FS model\cite{Paczynski1993,Meszaros1997,Sari1998} is not capable to account for the chromatic X-ray and optical temporal behaviour even if the jet structure\cite{Dai2001,Lipunov2001,Rossi2002} or the presence of the second wider jet\cite{Kumar2000a,RR2002,Peng2005} is taken into account. To resolve the chromaticity problem, we invoke the Large Angle Emission (also called high latitude emission, LAE, hereafter) from a structured jet, i.e. delayed prompt like emission from the jet envelope\cite{Oganesyan2020,Panaitescu2020}, which accounts for an additional X-ray component (see Methods).

\begin{figure}
    \centering
    \includegraphics[width=1.0\textwidth]{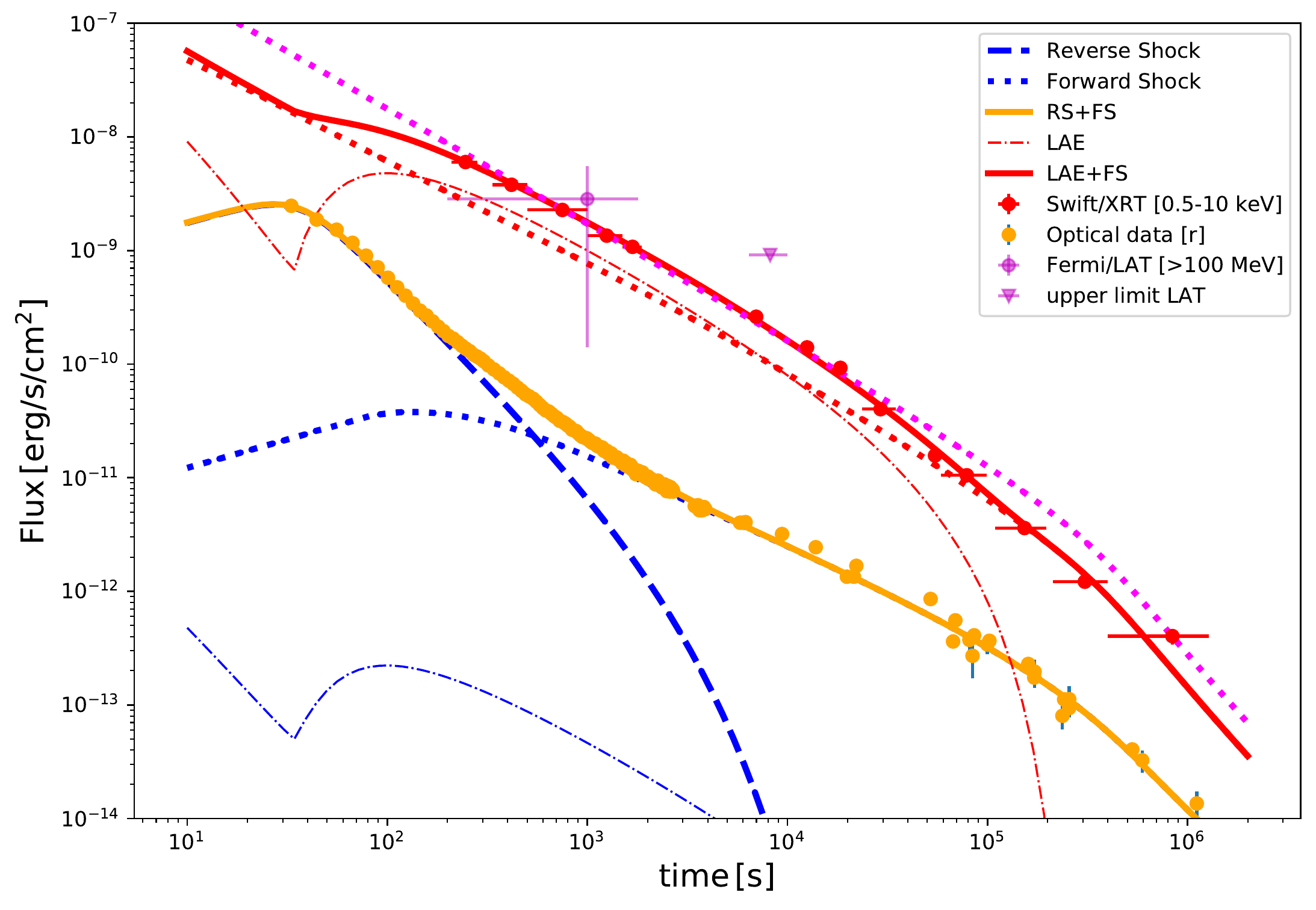}
    \caption{Optical (orange), X-ray (red) light curves together with the Fermi/LAT data points (magenta). We show the contributions from the reverse shock to the optical emission by dashed blue line and the forward shock emission by blue, red and magenta dotted line in the optical, X-ray and GeV bands, respectively. The Large Angle Emission is drawn by the dashed-dotted lines in the optical (blue) and X-ray (red) bands. The overall optical and X-ray models are shown by solid orange and red lines, respectively.}
    \label{fig:model}
\end{figure}

The joint optical/X-ray light curves are further modelled by a combination of the RS, FS and LAE components (see Fig.~\ref{fig:model}). To account in  
details of forward shock model we use the publicly available code {\it afterglowpy}\cite{Ryan2020}. 
We included empirically the RS component.
The best fit model is described by the RS component peaked at $\rm \sim 36$ s, with a temporal decay of $\rm \propto t^{-1.75}$ and a
characteristic time of $\rm t_{cut,RS} \sim 10^{3}$ s, after which 
the RS component drops exponentially. The cutoff-time corresponds to the passage of the cooling frequency through the observed optical band. This allows us to constrain the RS cooling frequency\cite{Kobayashi2000} at the shell crossing time $\rm \nu_{c,RS}(t_{0,RS}) \approx 3 \times 10^{17}$ Hz. We clearly see the expected softening of the optical spectrum from few hundreds of seconds until $\rm \sim 10^{3}$ s (Fig.~\ref{fig:lc}). The best fit parameters of the FS component suggest that the cooling synchrotron frequency of the FS at the RS crossing time is $\rm \nu_{c,FS}(t_{0,RS}) \sim 10^{18}$ Hz. We then 
carefully estimate the relative magnetisation parameter\cite{Zhang2003,Kobayashi2003} $\rm R_{B} = 2.5 \pm 0.3$ which gives the magnetic equipartition parameter of the RS component of $\rm \epsilon_{B,RS} = 0.12_{-0.10}^{+0.20}$. In principle, the RS can be formed in a magnetised ejecta\cite{Zhang2005,Giannios2008,Mizuno2009}. The FS model of GRB~210619B suggests that the circumburst medium is very rarefied, with a particle density of $\rm 6 \times 10^{-5} \, cm^{-3}$. The densities as low as $\rm 10^{-3}-10^{-5} \, cm^{-3}$ were inferred from the modelling of a few GRB afterglows\cite{Evans2014,Laskar2014,Laskar2015,Laskar2016,Alexander2017} and it could be connected with an unusual location of this GRB in the host galaxy or even be associated with an alternative progenitor such as a blue super-giant star \cite{Piro2014}. 

The low circumburst medium implies that the jet was initially propagating almost in a vacuum prior to the deceleration. Therefore, the ejecta has been spread and significantly decreased its magnetisation at the deceleration side providing the observed bright optical flash\cite{Granot2012}. This regime of formation of the RS emission (magnetised jet propagating into a vacuum\cite{Granot2012}) corresponds to $\rm  \epsilon_{B,RS} \sim T_{\gamma,1}/t_{RS} \sim 0.3$ roughly consistent with $\rm \epsilon_{B,RS} = 0.12_{-0.10}^{+0.20}$, where we have assumed that the RS ejecta corresponds to the duration of the first bright prompt emission pulse ($\rm \sim 10 \, s$). 
As predicted in this scenario, we observed equal deceleration times for the RS and FS.
Since the magnetisation of the RS ejecta has significantly decreased since times $t\rm \ll t_{RS}$\cite{Granot2012}, then 
the GRB jet could be initially highly magnetised, and can be even dominated by the Poynting flux\cite{Usov1992,Thompson1994,Meszaros1997b}.

Rich optical/X-ray/GeV data allow us for the first time to fully constrain the contributions from the RS, FS and the LAE components. We have demonstrated that one needs to invoke a narrow ($\rm 0.9_{-0.2}^{+0.1}$ deg) structured jet propagating into a very rarefied medium ($\rm \sim 6 \times 10^{-5} \, cm^{-3}$) to explain multi-wavelength observations of the GRB 210619B afterglow. While this GRB is produced by rare (but realistic) combination of observables,
it follows known spectral correlations\cite{Amati2002,Yonetoku2004,Ghirlanda2004}, and it shows a consistency with the standard internal shocks model for the production of the prompt 
emission (with an efficiency of $\rm 0.013_{-0.011}^{+0.020}$)\cite{Kobayashi1997,Daigne1998}. Furthermore, its true kinetic energy $\rm E_{kin,\theta} = 3.8_{-2.7}^{+10.2} \times 10^{52}$ erg is consistent with the energy of the population of supernovae associated with long GRBs\cite{Woosley2006}. An opening angle of $\rm \sim 1$ deg is rarely observed in GRBs\cite{Ryan2015} and is found in less than $\rm 10\%$ of studied populations\cite{Wang2018}, which are, however, usually embedded in environments with densities orders of magnitude larger. Extremely narrow opening angles $\rm <1.5^{o}$ and jets with steep structure $\rm k>4$ are expected in long GRBs due to the initial propagation of the jet into the progenitor material\cite{Salafia2020,Gottlieb2021}. However, the opening angles inferred from the observations of late-time afterglow steepening (jet breaks) suggest much wider jets\cite{Wang2018}. This discrepancy can be resolved by explaining the observed jet breaks as originated from viewing angle effects\cite{Kumar2003}. In this case, the "truly" on-axis GRBs are few and they expect to be the most luminous with the early jet break due to the narrow jet with a steep structure, exactly as witnessed in GRB 210619B. Importantly, the combination of observables suggests that the GRB jet could be initially magnetised\cite{Usov1992,Thompson1994,Meszaros1997b}. The fact that the GRB~210619B is located in a low-density medium and it is observed with the jet aligned towards us, allowed us to witness this rare and distant ($\rm z=1.937$) monster as source of an exceptionally bright optical transient ($\rm \sim 10$ mag).

\newpage

\newpage 

\newpage 

\newpage 

\begin{table}
\centering
\begin{tabular}{c|c|c|c}
\multicolumn{4}{c}{Mean Posterior Values} \\ \hline
&  \multirow{2}{*}{Parameters} & \multicolumn{2}{c}{Model} \\ \cline{3-4}
&  & RS+FS & RS+FS+LAE  \\ \cline{2-4} 
\multirow{8}{*}{Forward Shock} & $\rm E_{kin} $ & ($\rm 7.3 \pm 0.7) \times 10^{56} \, erg$ & $\rm 2.5_{-1.5}^{+10.1} \times 10^{56} \, erg$\\ 
& $\rm \theta_{j}$ & $\rm 0.2_{-0.01}^{+0.02} \, deg$ & $\rm 0.9_{-0.2}^{+0.1} \, deg$ \\ 
& $\rm k$ & $\rm 1.04_{-0.03}^{+0.07}$ & $\rm 6.2 \pm 2.3$ \\ 
& $\rm n $ & $\rm 2.00_{-0.75}^{+1.42} \times 10^{-8} \, cm^{-3}$ & $\rm 5.5_{-4.7}^{+8.6} \times 10^{-5} \, cm^{-3}$ \\
& $\rm p$ & $\rm 2.031_{-0.004}^{+0.003}$ & $\rm 2.023_{-0.008}^{+0.009}$\\
& $\rm \epsilon_{e}$ & $\rm 5.6_{-0.75}^{+0.91} \times 10^{-3}$ & $\rm 0.05_{-0.03}^{+0.05}$ \\
& $\rm \epsilon_{B}$ & $\rm 0.38_{-0.11}^{+0.13}$ & $\rm 7.9_{-7.7}^{+474} \times 10^{-4}$ \\ \hline
\multirow{3}{*}{Reverse Shock} & $\rm t_{RS}$ & $\rm 36 \pm 0.4 \, s$ & $\rm 36 \pm 0.4 \, s$ \\
& $\rm t_{RScut} $ & $\rm 1350_{-75}^{+85} \, s$ & $\rm 2290_{-215}^{+260} \, s$ \\
& $\rm \alpha_{RS}$ & $\rm -1.74 \pm 0.01$ & $\rm -1.75 \pm 0.01$ \\ \hline
\multirow{3}{*}{Large Angle Emission}  & $\rm R$ & & $\rm 7.9_{-1.3}^{+1.6} \times 10^{15} \, cm $\\
& $\rm \Gamma_{0}$ & & $\rm 360_{-250}^{+960}$ \\
& $\rm E_{p} $ & & $\rm 220_{-210}^{+5400} \, keV$ \\ \hline
\multirow{5}{*}{Inferred parameters} & $\rm E_{kin,\theta}$ & $\rm (4.4 \pm 0.3) \times 10^{51} \, erg$ & $\rm 3.8_{-2.7}^{+10.2} \times 10^{52} \, erg$ \\
& $\rm E_{\gamma,\theta} $ &$\rm 2.1_{-0.3}^{+0.4}\times 10^{49} \, erg$ & $\rm (4.4 \pm 1.4) \times 10^{50} \, erg$ \\
& $\rm \eta_{\gamma}$ &$\rm 4.7_{-0.4}^{+0.5} \times 10^{-3}$ & $\rm 1.3_{-1.1}^{+2.0} \times 10^{-2}$ \\
& $\rm R_{B}$ & $\rm 20 \pm 1$ & $\rm 2.5 \pm 0.3$ \\
& $\rm \epsilon_{B,RS}$ & $\rm 6_{-1}^{+3}$ & $\rm 0.12_{-0.10}^{+0.20}$ \\

\hline	

\end{tabular}
\caption{Mean posterior values of the parameters of two tested models (with and without the Large Angle Emission).}
\label{table:parameters}
\end{table}

\newpage 

\newpage

\clearpage


\clearpage

{\noindent  \bf Acknowledgements} 
GO is thankful to Om Sharan Salafia for the fruitful discussions.  SK and MP acknowledge support from the European Structural and Investment Fund and the Czech Ministry of Education, Youth and Sports \\
(Project CoGraDS -- CZ.02.1.01/0.0/0.0/15\_003/0000437).
FRAM-ORM operation is supported by the Czech Ministry of Education, Youth and Sports (projects LM2015046, LM2018105, LTT17006) and by European Structural and Investment Fund and the Czech Ministry of Education, Youth and Sports (projects CZ.02.1.01/0.0/0.0/16\_013/0001403 and CZ.02.1.01/0.0/0.0/18\_046/0016007).
The work is partially performed according to the Russian Government Program of Competitive Growth of Kazan Federal University.
This work was supported within the framework of the government
contract of the Special Astrophysical Observatory of the Russian
Academy of Sciences in its part entitled ``Conducting Fundamental
Research''. The research leading to these results has received funding from the European Union’s Horizon 2020 Programme under the AHEAD2020 project (grant agreement n. 871158). BB and MB acknowledge financial support from MIUR (PRIN 2017 grant 20179ZF5KS).

{\noindent  \bf Competing Interests} The authors declare no competing
interests


\clearpage

{\noindent \bf Materials and Methods}
\setcounter{figure}{0}
\setcounter{table}{0}
\renewcommand{\thefigure}{S\arabic{figure}}
\renewcommand{\thetable}{S\arabic{table}}

{\noindent \bf  Optical data} 

{\bf D50} \cite{d50} is a 50-cm telescope located at Ondřejov Observatory of Astronomical Institute of Czech Academy of Sciences, Czech Republic. It is equipped with Andor iXon Ultra 888 EMCCD camera and a set of photometric filters of the Sloan system. The telescope reacted to the trigger and started observing the position of the transient at 2021-06-19 23:59:53 UT ($T_0+28$ s, during the ongoing gamma-ray emission). In the following 2 hours, it acquired a series of unfiltered 10-s exposures, then followed by several longer exposures in the Sloan $g$, $r$, $i$ and $z$ filters at later times. The data have been processed and calibrated photometrically using PanSTARRS DR1 \cite{ps1} data of an ensemble of field stars, simultaneously deriving both the zero point and color term for unfiltered images. The best fit for the instrumental photometric system is $\mbox{mag} = r - 0.05(g - r) + \mbox{ZP}$, 
very close to PanSTARRS $r$.

{\bf FRAM-ORM} \cite{fram_la_palma} is a 25 cm f/6.3 telescope located at Observatorio del Roque de los Muchachos, La Palma, Spain. The telescope is equipped with B, V, R and z filters, and a custom Moravian Instruments G2-1000BI camera based on a back-illuminated CCD47-40 chip. It started observing the transient position at $T_0+30$ s and acquired a series of unfiltered 20-s exposures until $T_0+1050$ s. Due to strong wind at the telescope site the images were distorted. In order to exclude the possible varying influence of nearby stars on the photometric measurements, the frames have been passed through a difference imaging pipeline based on HOTPANTS image subtraction code \cite{hotpants}. The measurements have been calibrated using field stars and PanSTARRS DR1 catalogue \cite{ps1}. The best fit for the instrumental photometric system is $\mbox{mag} = r + 0.10(g - r) + \mbox{ZP}$, very close to PanSTARRS $r$, although slightly different from the system of D50.

{\bf Mini-MegaTORTORA} (MMT-9) nine-channel wide-field monitoring system \cite{beskin_mmt_2017} is a set of 9 7-cm objectives equipped with Andor Neo sCMOS cameras and independently installable Johnson-Cousins B, V, R and polarimetric filters, and mounted on 5 mounts. The system responded to the BAT trigger and observed the position
of the transient since 2021-06-20 00:00:20 UT
($T_0$+55 s) and until 2021-06-20
00:10:28 UT ($T_0$ + 663 s). The system simultaneously acquired series of
frames with 1 s exposures, 5 s exposures and 30 s exposures in white light,
10 s exposures in B filter, and 10 s exposures in V filter. 
The transient is clearly detectable in all the acquired sequences except in the B filter ones. As the field was crowded due to the large pixel scale of Mini-MegaTORTORA, the photometry has been performed on 
the images after the subtraction of a template image using the HOTPANTS image subtraction code \cite{hotpants}. The template has been constructed by averaging a large number of frames from observations of the same position on successive nights.
Photometric measurements in Johnson B and V filters have been calibrated to Vega magnitudes using a large ensemble of field stars and their photometric measurements in Gaia DR2 catalogue 
converted to Johnson-Cousins system using conversion formulae based on Stetson standards.
Also, both white light images and the ones in V filters have been calibrated to AB magnitudes using PanSTARRS DR1 catalogue \cite{ps1} data of field stars by fitting the color terms and zero points for every sequence. The best fit for the unfiltered photometric system is $\mbox{mag} = r + 0.353(g - r) + \mbox{ZP}$, while for the V filter 
is $\mbox{mag} = r + 0.538(g - r) + \mbox{ZP}$. 

\begin{figure}[ht!]
    \centering
    \includegraphics[width=0.9\textwidth]{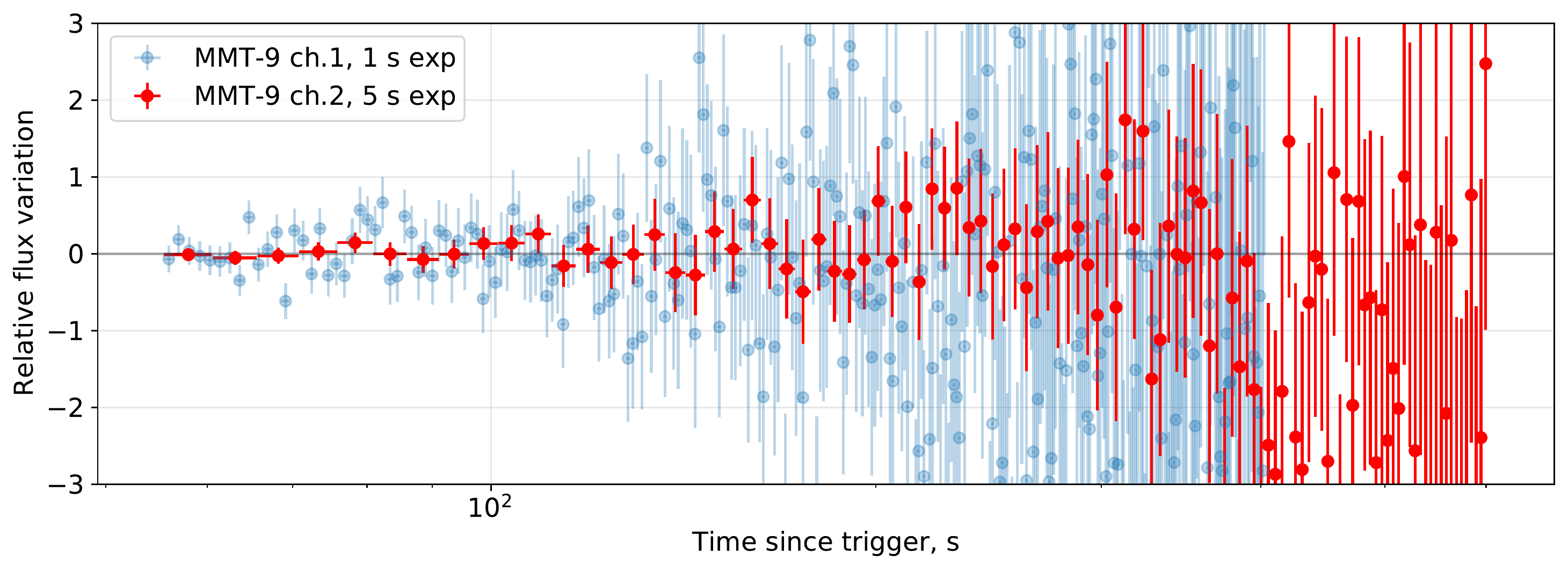}
    \caption{Relative flux variations of the high temporal resolution (1 s and 5 s exposures) data from Mini-MegaTORTORA. The flux is normalized to the smooth curve corresponding to D50 data.}
    \label{fig:relflux_variation}
\end{figure}

{\bf High temporal resolution} Mini-MegaTORTORA data are acquired from the channels 1 and 2 that were operating with 1 and 5 s exposures. Their light curves are generally similar to the D50 one acquired with 10 s temporal resolution, and do not show any significant excess variability relative to its smooth behaviour (see Fig.~\ref{fig:relflux_variation}), despite still ongoing and highly irregular gamma-ray activity. This excludes that optical and prompt gamma-ray emissions are produced by the same process, and it is consistent with optical emission being mostly produced by a reverse shock.

{\bf Early optical color} has been estimated using simultaneous observations in slightly different photometric systems. If both systems are represented as $r + C(g-r)$, then the ratio of fluxes among them is $F_1/F_2 = 10^{-0.4*(C_1-C_2)(g-r)}$. We computed the average ratios of fluxes from all early data sets (unfiltered FRAM-ORM, unfiltered Mini-MegaTORTORA, V filter Mini-MegaTORTORA) relative to the interpolated flux from D50 in several temporal intervals. We then fitted them to evaluate the color index. The results are shown in Figure~\ref{fig:lc} and Table~\ref{table:colors}.

{\bf Galactic extinction} towards the transient has been taken from NASA/IPAC Extragalactic Database (NED) Extinction Calculator, and then used to correct the optical fluxes used for plotting Fig.~\ref{fig:lc}. It corresponds to $A_{\rm V}^G=0.467$ and $E_{g-r}^G=0.17$. 
Later time spectral\cite{gcn_redshift}
and multicolor\cite{gcn_30271} data suggest the presence of an additional reddening of the afterglow, 
shown as a strong broad absorption feature at $\sim$4500~AA supposedly corresponding to a 2175~AA dust feature due to an intervening absorber at $z=1.095$. 
To model it, we used the data\cite{gcn_30271} in $g$, $r$, $i$ and $z$ filters measured at around $T_0+10000$ s, and we assumed a power-law optical spectrum and a Milky Way like absorption law\cite{pei92} with $A_{\rm V}=0.30\pm0.05$. This results in $E_{g-r}=0.40$. The fit required no additional reddening due to the burst host galaxy \textit{per se}, so we assume that it is negligible.

{\bf }

{\noindent \bf  X-ray data}
We have retrieved the X-ray (0.3-10 keV) lightcurve of GRB 210619B from the {\it Swift} Science Data Center supported by the University of Leicester\cite{Evans2009}. Four time-bins in the Window Timing mode (335-1864 s) and ten time-bins in the Photon Counting mode (209-283 s ,6395-$\rm 1.3\times10^{6}$ s) were selected for the spectral analysis. The source and background spectral files, the ancillary response and the response matrix files were produced by the Online {\it Swift}-XRT GRB spectrum repository\cite{Evans2009}. We have fitted all 14 unbinned XRT spectra by a simple power-law model using XSPEC (12.10.1) and 
adopting C statistic likelihood. To account for the Galactic\cite{Kalberla2005} and intrinsic metal absoption we applied {\it tbabs} and {\it ztbabs} models, respectively. The hydrogen column of the host galaxy $\rm N_{H}$ has been left as a free parameter. We do not find significant variations of $\rm N_{H}$ among 14 spectral fits. Finally, we have estimated unabsorbed X-ray flux in the 0.5-10 keV band and used it for modelling the joint X-ray/optical afterglow lightcurves. We report the flux of all 14 time-bin in the Table \ref{xrt_tab} in Supplementary Materials.   

{\noindent \bf  \textit{Fermi}/GBM data}
We have extracted the time-resolved spectra of \textit{Fermi}/GBM corresponding to the time-windows of D50 and MMT-9 observation times. \textit{Fermi}/GBM is composed by 12 sodium iodide (NaI) and two bismuth germanate (BGO) scintillation detectors\cite{Meegan2009}. For each time-bin we have chosen spectral data from two NaI detectors and one from a BGO detector. The choice of NaI and BGO detectors is based on the source position angle. We have chosen the NaI-4 (51 deg), NaI-8 (26 deg) and BGO-1 (76 deg) detectors. We have extracted the Fermi/GBM spectra by the GTBURST tool. To model the background, we have fitted pre- and post-burst data by the energy- and time-dependent polynomial. The time-resolved spectra have been fitted by Band function\cite{Band1993} using XSPEC (12.10.1) and applying the PGSTAT likelihood. We took the spectral data of NaI from 8 keV to 900 keV, and the BGO data from 300 keV to 10 MeV. The best fit parameters are listed in the Table \ref{fermi_tab} in Supplementary Materials. We report the low energy photon index $\rm \alpha$, the high energy photon index $\rm \beta$, the break energy $\rm E_{0}$ and the observed flux in the range between 8 keV and 10 MeV. 
In order to estimate the isotropic equivalent energy $\rm E_{iso}$ and the peak luminosity $\rm L_{p,iso}$ of GRB 210619B, we also model with the Band function the time-averaged spectrum and one-second peak spectrum. The time-integrated spectrum is defined as of the duration of the GRB ($\rm T_{90}=54.785$ s from 0.576 s to 55.361 s as defined by the Fermi/GBM Burst catalog). The one-second peak time spectrum is defined between 0.51 and 1.51 s as reported by Fermi/GBM team\cite{gcn_fermi}. Due to the extreme brightness of the spectrum, we notice the presence of the Iodine K-edge at 33.17 keV\cite{Meegan2009} and therefore we ignore the NaI data from 30 keV to 40 keV for the time-integrated and one-second peak time spectrum. We also extend the spectral coverage by BGO up to 40 MeV. 

\begin{figure}[ht!]
    \centering
    \includegraphics[width=1.0\textwidth]{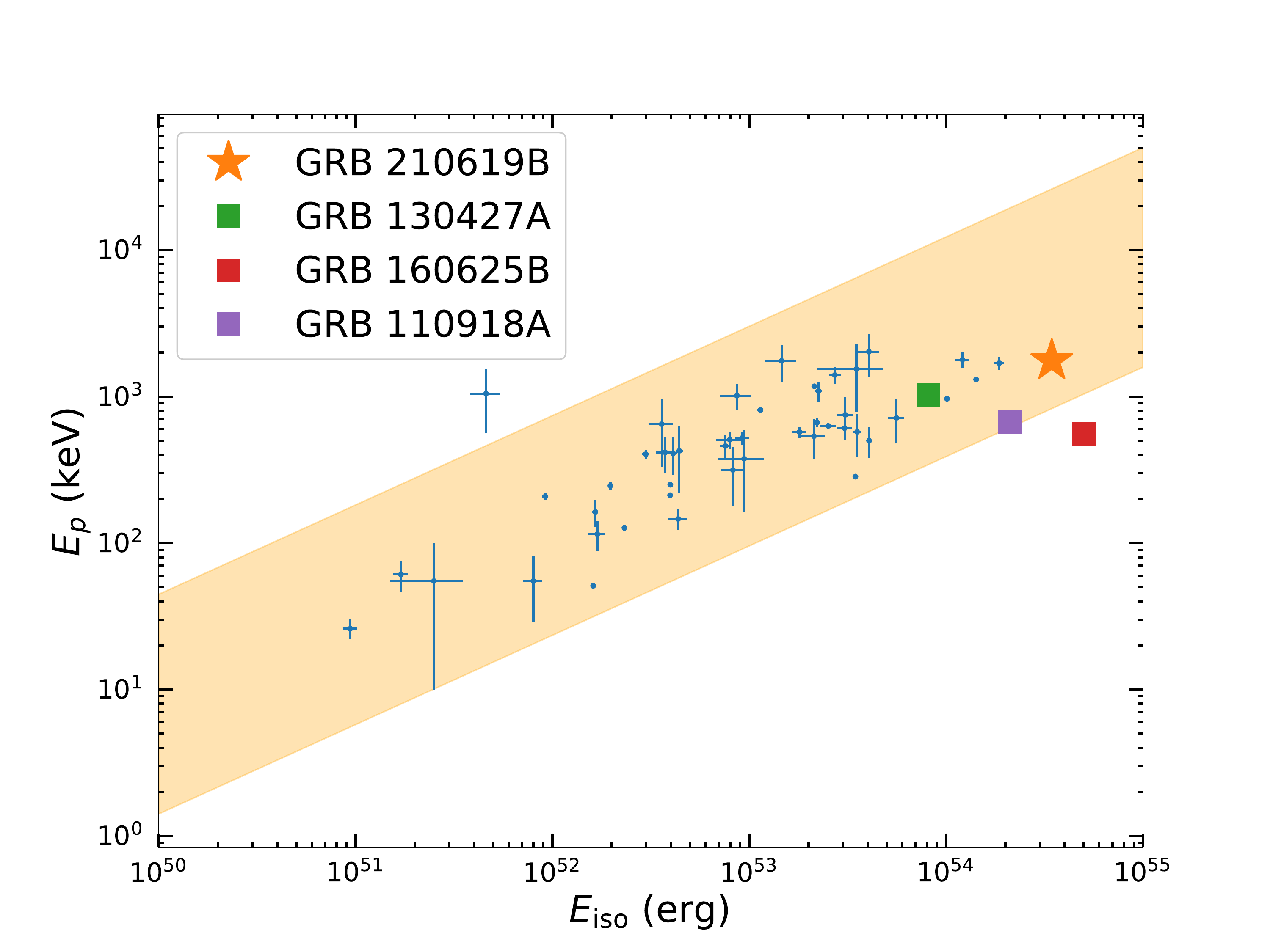}
    \caption{Amati relation for GRB samples with known redshift and estimated $E_{peak}$. The quantities $E_{p}$ and $E_{iso}$ represent the rest frame peak energy and the isotropic energy, respectively. 
    The shaded region marks the 3$\sigma$ scatter of the point distribution
around the best fit line \cite{2012MNRAS.421.1256N}. The blue markers represent the 46 GRBs detected with known $E_p$ and redshift and retrieved from \cite{2012MNRAS.421.1256N} and references therein.
GRB 210619B (this work) belongs to the 3$\sigma$ contour around the Amati correlation, and is one of the brightest sources.
GRB 130427A\cite{2014Sci...343...48M} is marked with a green square which showed largest isotropic energy releases. 
The historical outlier, GRB 160625B\cite{2018NatAs...2...69Z}, is shown with a red square.
GRB 110918A\cite{2013ApJ...779..151F} with the highest isotropic-equivalent luminosity observed so far, is marked with a purple square marker. }
    \label{fig:Amati}
\end{figure}

{\noindent \bf  GRB~210619B and spectral-energy correlations}
The best fit parameters of the time-integrated spectrum are the following: the low-energy photon index $\rm \alpha=-0.86 \pm 0.02$, the high-energy photon index $\rm \beta=-2.18_{-0.03}^{+0.02}$, the break energy in the photon spectrum 
$\rm E_{0}=178.67_{-6.84}^{+7.16}$ keV with PGSTAT equal to 788 for 339 degrees of freedom. 
The flux of $\rm T_{90}$ spectrum is estimated to be $\rm (6.67 \pm 0.08) \times 10^{-6} erg \, cm^{-2} \, s^{-1}$ between 0.34 keV and 3404.8 MeV. This band is used to compute isotropic equivalent energy $\rm E_{iso}$ and the isotropic equivalent peak luminosity $\rm L_{p,iso}$ in the energy band 1 keV - 10 MeV in the rest frame given the redshift of 1.937. 
We estimate $\rm E_{iso}=(3.44 \pm 0.04) \times 10^{54} \, erg$, and the peak energy of the energy flux spectrum in the rest frame $\rm E_{peak}=598_{-23}^{+24} \, keV$. The total energy and the spectral peak are in agreement with the Amati relation\cite{Amati2002}(see Fig.~\ref{fig:Amati}). GRB~210619B is the second most energetic GRB after GRB 160625B. 

\begin{figure}[ht!]
    \centering
    \includegraphics[width=1.0\textwidth]{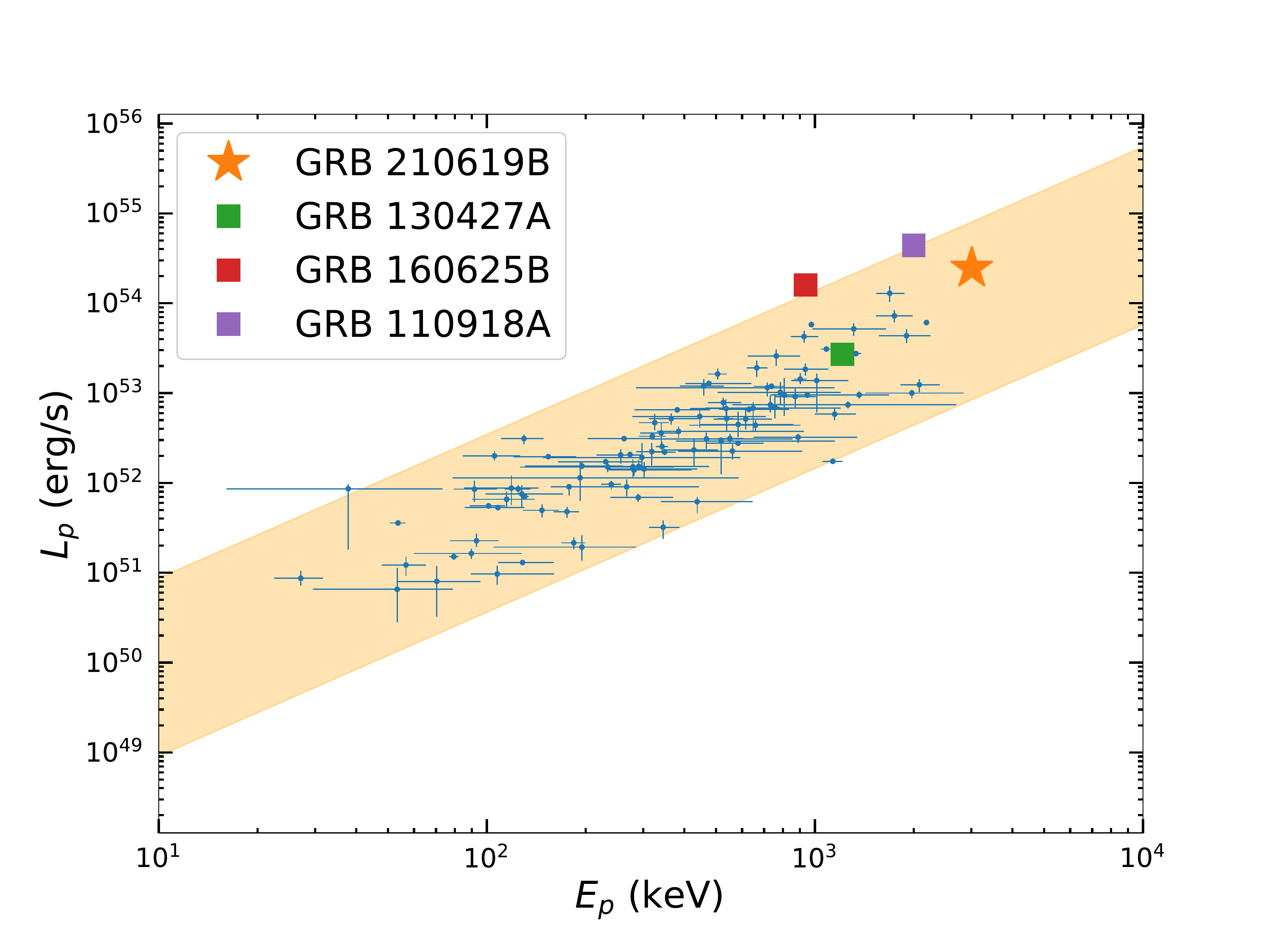}
    \caption{
    Yonetoku relation, where quantities $L_{p}$ and $E_{p}$ represent the 1-second peak luminosity and the  rest frame peak energy, respectively. 
    The data set\cite{2010PASJ...62.1495Y} (blue markers) is a collection of short GRBs, long GRBs and GRBs with high redshift (z$>$6).
    The shaded region marks the 3$\sigma$ scatter of the point distribution
around the best fit line \cite{2010PASJ...62.1495Y}. 
GRB 210619B (this work) belongs to the 3$\sigma$ contour around the Yonetoku correlation.
The GRBs, namely GRB 130427A\cite{2014Sci...343...48M}, GRB 160625B\cite{2018NatAs...2...69Z}, GRB 110918A\cite{2013ApJ...779..151F} are marked with
green, red and purple square markers, respectively.
    }
    \label{fig:Yonetoku}
\end{figure}

The best fit parameters of the one-second peak time spectrum are the following: the low-energy photon index $\rm \alpha=-0.36 \pm 0.03$, the high-energy photon index $\rm \beta=-2.16 \pm 0.03$, the break energy in the photon spectrum 
$\rm E_{0}=212.88_{-11.35}^{+11.91}$ keV with PGSTAT equal to 548 for 339 degrees of freedom. We estimate the flux of one second spectrum $\rm (8.8 \pm 0.1) \times 10^{-5} erg \, cm^{-2} \, s^{-1}$ between 0.34 keV and 3404.8 MeV. The corresponding isotropic equivalent peak luminosity in the rest frame is $\rm L_{p,iso}=(2.45 \pm 0.03) \times 10^{54} \, erg \, s^{-1}$ and the peak energy of the energy flux spectrum in the rest frame is $\rm E_{peak}= 1024_{-55}^{+57} \, keV$. The peak luminosity and the corresponding peak energy are consistent with the Yonetoku relation\cite{Yonetoku2004}(see Fig.~\ref{fig:Yonetoku}). GRB~210619B is positioned as the second most luminous GRB after GRB~110918A and it has the largest spectral peak energy among long GRBs in the Yonetoku plane.

{\noindent \bf  {\it Fermi}/LAT} \\
\\
The Large area Telescope (LAT) on board \textit{Fermi} consists of a tracker and a calorimeter sensitive to the gamma-ray photons of energy
between 30 MeV and 300 GeV \cite{2013ApJS..209...11A}. 
We used GTBURST tool from the official \textit{Fermi}-software to extract and analyse the data. 
Due to its position (R.A. $= 319.71^{\circ}$ and Dec. $= 33.86^{\circ}$), the source was out of the field of view (FoV) of LAT during the first 200 seconds after trigger.
The LAT data were analyzed in two periods: a) $T_0 + 200$\,s to $T_0 + 1800$\,s, and b) $T_0 + 6400$\,s to $T_0 + 10000$\,s in the energy band of above 100\,MeV.
For this analysis, we used a region of interest (ROI) of 12$^{\circ}$
centred at the burst position provided by {\it Swift}/XRT \cite{gcn_30270}. 
In addition, a zenith angle cut of 100$^{\circ}$ was applied to reduce the contamination of the gamma-ray photons from the Earth limb. 
We assumed a spectral model of type "powerlaw2" for the source.
We considered "isotr template" and "template (fixed norm.)" 
for the particle background and the Galactic component, respectively.
The calculations of the flux in the energy bin given above are calculated with the "unbinned likelihood analysis" considering a minimum test statistics (TS$_{\rm min}$) of 10.
The \textit{Fermi}/LAT analysis shows that for period a) the source is found with an energy flux of (2.84$\pm$2.70)$\times$10$^{-9}$ erg s$^{-1}$ cm$^{-2}$ and with a TS value of 12. The spectral index for this period is -1.68$\pm$0.32. For period b), the TS of the source is 8 and the flux upper limit is 9.13$\times$10$^{-10}$ erg s$^{-1}$ cm$^{-2}$.

\begin{figure}[ht!]
    \centering
    \includegraphics[width=1.0\textwidth]{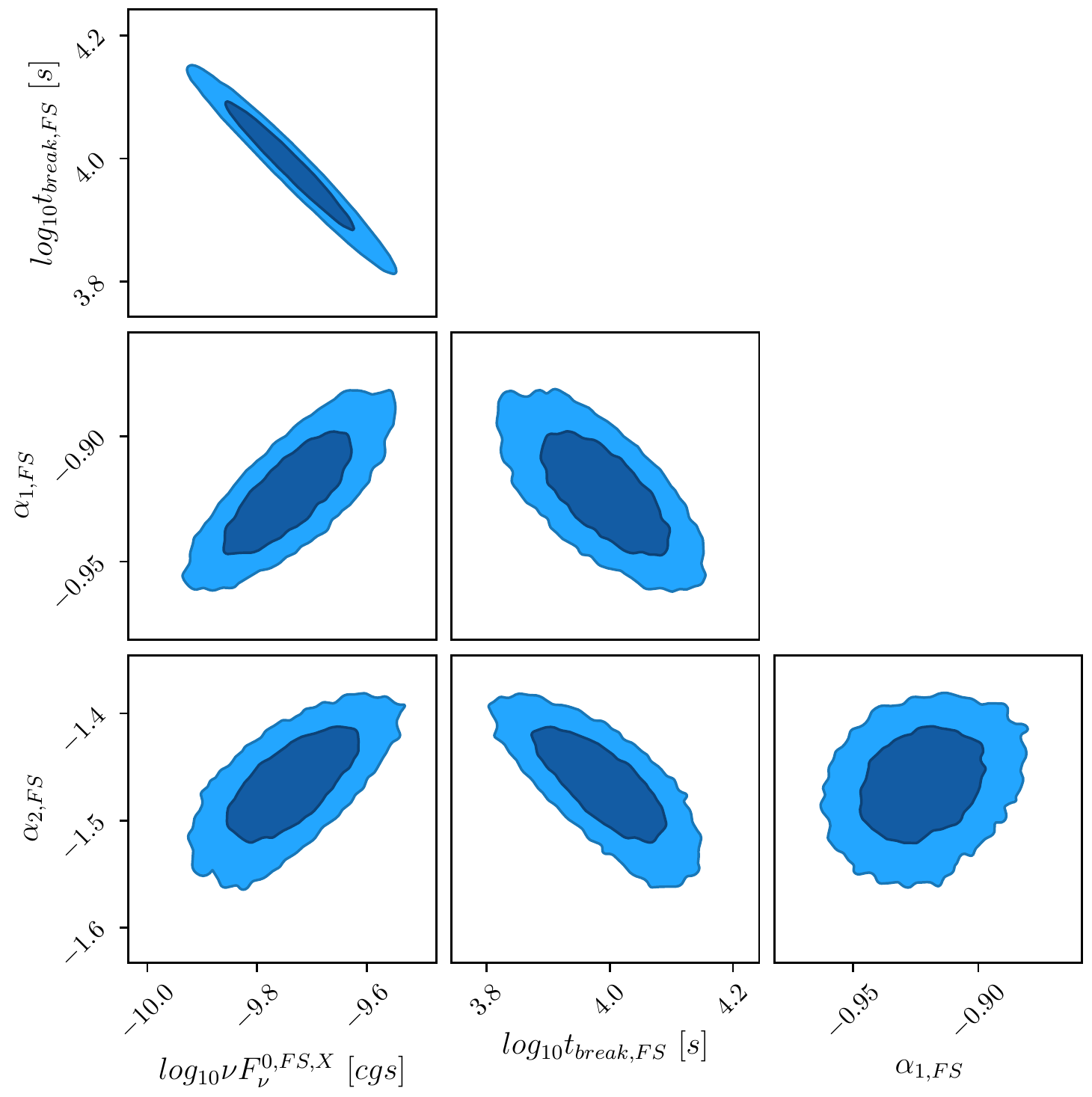}
    \caption{The corner plot for the parameters of the empirical model for the X-ray light curve.
    }
    \label{X-ray_empirical}
\end{figure}

{\noindent \bf The empirical model for the X-ray/optical light curves}
We first model the X-ray (0.5-10 keV) and optical lightcurves  empirically in order to assess the temporal evolution of the
possible reverse and forward shock components. The X-ray lightcurve is modelled by a smoothly connected broken power-law function 
\begin{equation}\label{empiric}
\nu F_{\nu, \text {X}}(t)=F_{0}^{0,FS,X}\left[\frac{1}{2}\left(\frac{t}{t_{break,FS,X}}\right)^{-n \alpha_{1}}+\frac{1}{2}\left(\frac{t}{t_{break,FS,X}}\right)^{-n \alpha_{2}}\right]^{-(1 / n)}
\end{equation}
which returns an initial decay of the energy flux $\rm \nu F_{\nu}^{X} \propto t^{-0.9}$ before the temporal break of $\rm t_{break,FS,X} \sim 10^{4}$ s. This is then followed by a flux decay $\rm \nu F_{\nu,X} \propto t^{-1.5}$ after the temporal break.
We have adopted an intermediate smoothness parameter $n=5$. The spectral index in 0.5-10 keV remains roughly the same $\rm \beta_{X} \approx 0.9$ (where $\rm F_{\nu,X} \propto \nu^{-\beta_{X}}$) before and after the temporal break (see Fig.~\ref{X-ray_empirical} for the distribution of the parameters).

\begin{figure}[ht!]
    \centering
    \includegraphics[width=1.0\textwidth]{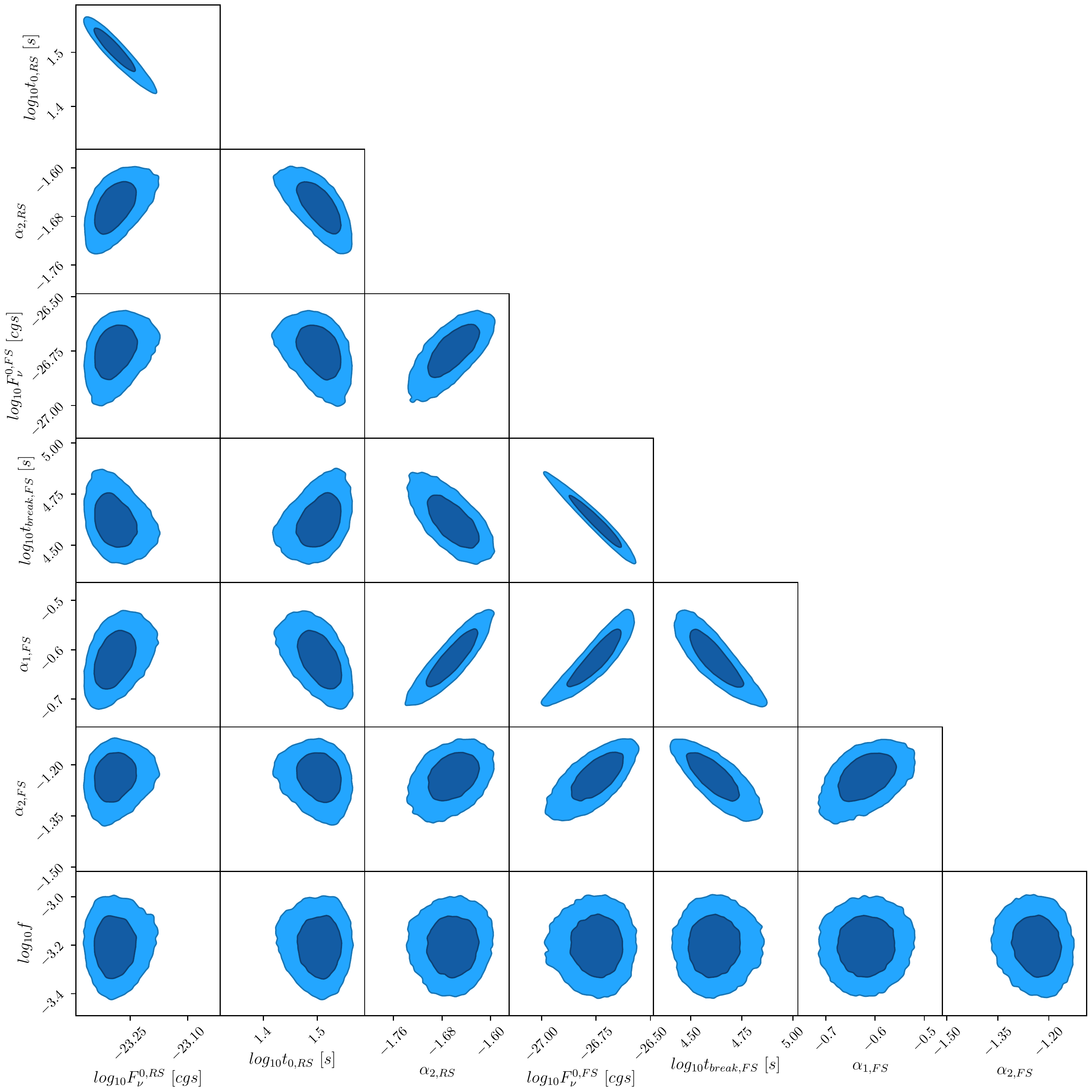}
    \caption{The corner plot for the parameters of the empirical model for the optical light curve.
    }
    \label{optical_empirical}
\end{figure}

To account for the complex temporal structure of the optical lightcurve, we adopt two smoothly broken power-law functions, one representing the reverse shock component (RS hereafter) and the other the forward shock component (FS hereafter). Since we do not observe the rise of the early optical emission, we fix the initial rise index of the RS component to the theoretically driven value of $\alpha_{1,RS}=0.5$\cite{Nakar2004}, where $\rm F_{\nu,RS,opt} \propto t^{\alpha_{1,RS}}$. To account for the adiabatic cooling of the ejecta\cite{Kobayashi2000} which has previously experienced the passage of the RS, we introduce an exponential cutoff in the empirical function of the RS at $t_{cut,RS}$. The candidate FS component is modelled as in the case of X-ray lightcurve, with the only difference that we use $\rm F_{\nu,opt}$ data as input for the  empirical function \ref{empiric}. The model returns a peak flux density of the RS component of $\rm F_{\nu}^{0,RS} \sim 0.5 Jy$ and a decay index of $\rm \alpha_{2,RS}=-1.6$, a pre-break decay index of the candidate FS component of $\rm \alpha_{1,FS,opt} \approx -0.6$, a temporal break of $\rm t_{break,FS,opt} \approx 4 \times 10^{4} s$ and a post-break index of $\rm \alpha_{2,FS,opt} \approx -1.2$. The cutoff time $t_{cut,RS}$ of the RS component is not 
constrained in the fitting by the empirical model (see Fig.~\ref{optical_empirical} for the distribution of the parameters).

The spectral index in the R-band is derived by the color differences throughout all the observed emission. It starts (at $\rm \sim 30 s$) with a quite low value of $\beta_{opt} =-2.10$ (however with huge uncertainty of $2.43$), then it fluctuates around $\rm \beta_{opt} \approx 1.0$ rising towards softer spectral values until the end of the RS component. With the rise of the candidate FS component the optical spectral index shows a softening trend from  a value of $\beta_{opt} = 0.37 \pm 0.26 $ to $\beta_{opt} = 1.40 \pm 0.32 $. Since the redenning from the host galaxy is unknown, the absolute values $\beta_{opt}$ is uncertain but the overall softening of the spectra during the RS and FS components is expected to remain the same, since the additional correction by the host would affect all the values by the same $\delta \beta_{opt}$.
It is remarkable the overall consistency of the spectral indices derived accurately in the X-ray band and by the color differences in the optical bands, except that of $\rm \approx 1000 \, s$, at the time of  transition between RS and the candidate FS components. 

We then compare the spectral shapes and the temporal behaviour of optical and X-ray emission with the expectations from the standard afterglow model of GRBs\cite{Paczynski1993,Meszaros1997,Sari1998}. The standard FS model is based on the ultra-relativistic, adiabatic, self-similar spherical solution of Blandford $\&$ McKee for a dynamics of a point like explosion\cite{BM1976}. The radiation is computed assuming a synchrotron emission from shocked-accelerated non-thermal electrons. For our analytical estimates, we ignore the effects of the synchrotron self Compton (SSC) cooling of electrons. This system is characterised by the following usual parameters: $\rm E_{kin}$ the kinetic energy of the spherical ejecta, $\rm n$ the density of the circumburst medium, $\rm \epsilon_{e}$ the fraction of kinetic energy given by non-thermal electrons, $\rm \epsilon_{B}$ the fraction of the kinetic energy in the magnetic field, $\rm p$ the spectral index of power-law distributed electrons, and $\rm \theta_{j}$ the opening angle of the jet. We assume that all electrons that received $\rm \epsilon_{e} E_{kin}$ got accelerated. We also assume an homogeneous circumburst medium. We further argue in terms of the synchrotron $\rm \nu_{m}$ and cooling $\rm \nu_{c}$ frequencies, which are the observed frequencies corresponding to the minimum Lorentz factor of accelerated electrons and the cooling Lorentz factor of electrons, respectively. We further compare the observed temporal and spectral indices with expectations from the FS model\cite{Granot2002,Gao2013}.

We assume that before the $\rm t_{break,FS,X}$ the X-ray band is located between $\rm \nu_{m}$ and $\rm \nu_{c}$. Considering an homogeneous circumburst medium, in order to produce the initial decay of $\rm \nu F_{\nu}^{X} \propto t^{-0.9}$, one would require an electron index of $\rm p \approx 2.2$ which would imply an XRT spectral index of $\rm \beta_{X} \approx 0.6$ harder than observed $\rm \beta_{X,obs} \approx 0.9$. In our assumed case, i.e. when $\rm \nu_{m}<\nu_{X}<\nu_{c}$ at $\rm t<t_{break,FS,X}$, the temporal break $\rm t_{break,FS,X}$ could be caused by the passage of 
$\rm \nu_{c}$ through the XRT energy band at $\rm t_{break,FS,X}$. However, it is highly inconsistent with the observed temporal break of the FS component in the optical light curve $\rm t_{break,FS,opt} \approx 4 \times 10^{4} $ s, since the expected time of this spectral transition in the R band is $\rm t_{break,\nu_{c},opt} \geqslant t_{break,FS,X} (\frac{\nu_{X}(t_{break,FS,X})}{\nu_{R}})^{2} \geqslant 2 \times 10^{7} \, s$. Therefore, we can exclude the scenario when the temporal optical/X-ray break is associated with the cooling frequency. 

We are then left with another interpretation of the temporal break, which is an early jet break\cite{Rhoads1997,Sari1999b} at $\rm \sim 10^{4} \, s$. To identify the temporal shape of the light curve after $\rm t_{break,FS,X} \approx t_{break,FS,opt}$, we first assume that $\rm \nu_{m}<\nu_{opt,X}<\nu_{c}$
at $\rm \sim 10^{4} \, s$. This is justified by the rough consistency of the observed optical and X-ray spectral indices $\rm \beta_{X} \approx \beta_{opt}$. For the previously inferred $\rm p \approx 2.2$, the temporal index after the $\rm \sim 10^{4} \, s$ should be $\rm \alpha_{X}=\alpha_{opt}=3p/4 \approx 1.65$ steeper than observed $\rm \nu F_{\nu,X} \propto t^{-1.5}$. However, if we assume that $\rm \nu_{m}<\nu_{R}<\nu_{c}$ while $\rm \nu_{X}>\nu_{c}$, we would require very soft $\rm p \approx 1.9$. And if we insist on $\rm p \approx 2$, the temporal index after the jet break would be even steeper $\rm \alpha_{X} \approx 1.75$. Therefore, the most optimal scenario would correspond to the case of $\rm \nu_{X} \leqslant \nu_{c}$, i.e. the cooling frequency is slightly above the XRT energy range. This is additionally supported by the quite hard value of the spectral index measured in the Fermi/LAT band (30 MeV-300 GeV) $\rm \beta_{LAT} = 0.68 \pm 0.32$ which corresponds either to the very end of the synchrotron spectrum or to the initial rise of the SSC component. While the most optimal scenario ($\rm \nu_{X} \leqslant \nu_{c}$) is roughly consistent 
with the temporal shape of the X-ray light curve, it is highly inconsistent with the joint chromatic optical light curve. The optical light curve is shallower than the X-ray light curve before the temporal break ($\rm \alpha_{1,FS,opt} \approx -0.6$ vs $\rm \alpha_{1,FS,X} \approx -0.9$) and after ($\rm \alpha_{2,FS,opt} \approx -1.2$ vs $\rm \alpha_{2,FS,X} \approx -1.5$). Additionally, the observed temporal breaks are not simultaneous as expected from the jet break and the X-ray spectra are much softer than expected in $\rm \nu_{m}<\nu_{opt,X}<\nu_{c}$. Therefore, one needs to invoke an additional process/component in order to describe both optical and X-ray light curves. We also notice that the wind medium would return more problematic solutions for the FS component, since even for $\rm \nu_{m}<\nu_{opt,X}<\nu_{c}$, very soft electron spectrum $\rm p \approx 1.5$ would be required. 

{\noindent \bf X-ray/optical light curves from external shocks}
We model the joint optical and the X-ray light curves with the FS model using publicly available code {\it afterglowpy}\cite{Ryan2020}. The adopted model corresponds to a Power law structured jet (without the spreading effects) decelerating in a homogeneous circumburst medium. We have selected a power-law structured jet because we expect to see the difference in the shape of the light curve compared to the top-hat jet at late times if the early temporal break at $\rm \sim 10^{4}$ s is associated with the jet break. We set the viewing angle of the observer to zero, since we observe one of the most luminous and energetic GRB. We include the synchrotron and SSC radiation components. To account for the RS component, we adopt the previously described phenomenological function with fixed rising shape and the smoothness parameter. We exclude the fractional parameter $\chi$ (fraction of accelerated parameters) since we found that it is not constrained and it does not change the typical values of other parameters. 

\begin{figure}[ht!]
    \centering
    \includegraphics[width=1.0\textwidth]{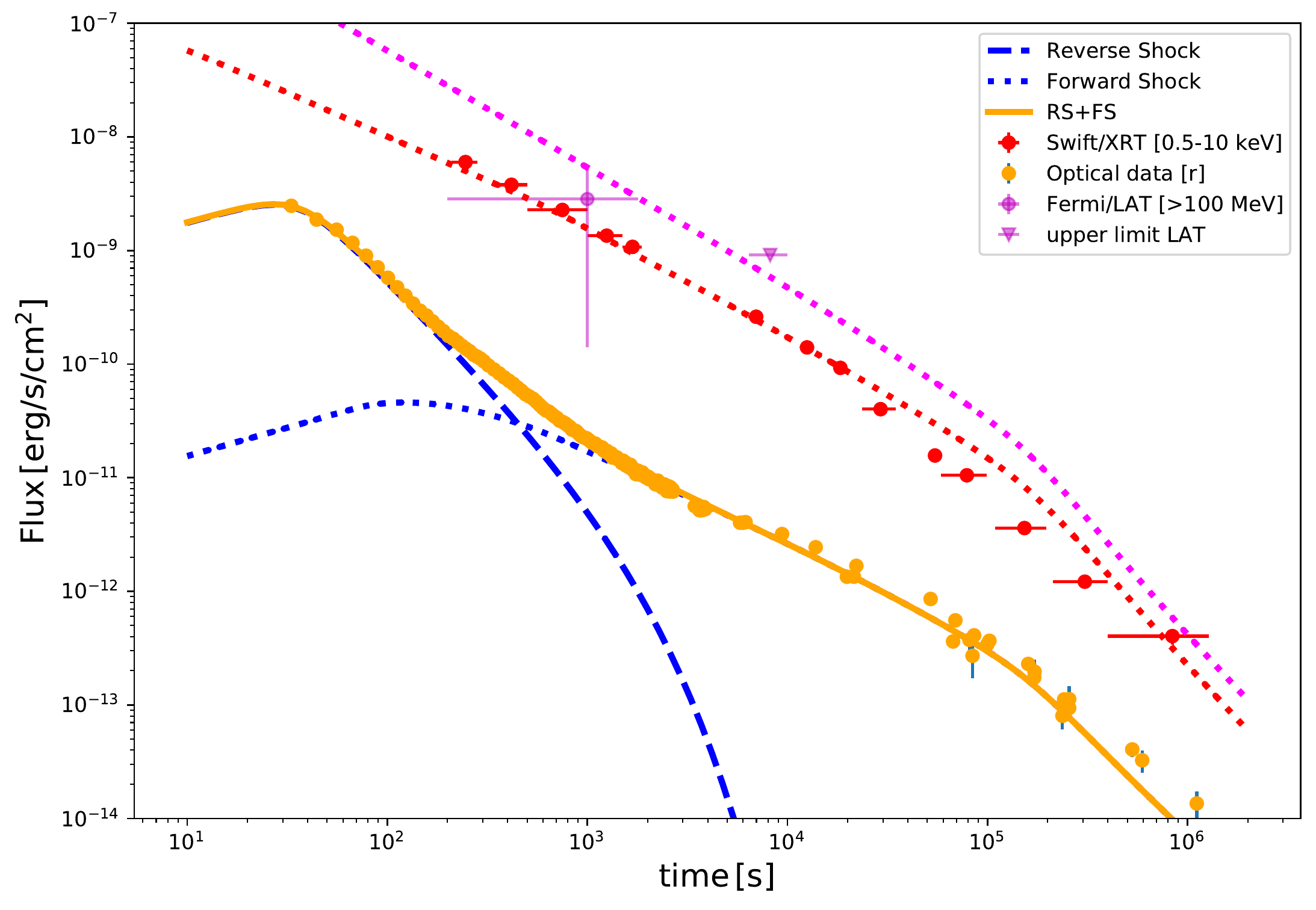}
    \caption{The FS+RS model and the optical/X-ray/LAT data. 
    }
    \label{the_model}
\end{figure}

The modelling returns a jet with an isotropic equivalent of kinetic energy of $\rm E_{kin} \sim 7 \times 10^{56}$ erg and the opening angle of $\rm \sim 0.2\ deg$, propagating in a 
rarefied circumburst medium with $\rm n \sim 2 \times 10^{-8} \, cm^{-3}$. The microphysical parameters are the following: $\rm \epsilon_{e} \sim 0.006$, $\rm \epsilon_{B} \sim 0.4$ and $\rm p \approx 2.03$. In this joint X-ray/optical modelling we 
were able to better constrain  the parameters of the RS component: $\rm t_{0,RS} \sim 36$s, $\rm F_{\nu}^{0,RS} \sim 0.5$ Jy, $\rm \alpha_{2,RS} \approx -1.7$, $t_{cut,RS} \approx 1.4\times 10^{3}$s. These parameters are obtained by considering the optical data corrected for the Galactic absorption and the absorber at z=1.1, but it does not take into account the unknown reddening by the host galaxy. The combined FS and the RS models together with the optical/X-ray data are shown in Fig.~\ref{the_model}. One can notice that the FS model is not able to reproduce the observed chromatic light curves, as expected from the previous analytical estimates. We have additionally tried to model the light curves by the power-law structured jet, two-component jets and we have failed to reproduce the fast declining X-ray light curve together with chromatic shallower optical light curve. One could assume an additional energy injection to the decelerating forward shock\cite{Dai1998,Zhang2001} to flatten the optical light curve, however this would necessarily produce similar temporal behaviour also in X-rays\cite{Fan2006} which is not observed. 

{\noindent \bf X-ray/optical light curves from external shocks and the Large Angle Emission}
To account for the chromatic X-ray and optical temporal behaviour, we have included the Large Angle Emission (LAE)\cite{Fenimore1996,Kumar2000}, i.e. the delayed prompt like emission received from the higher latitudes of the jet. It was suggested that the LAE from structured jets could account for the diverse X-ray afterglows observed in GRBs, including plateaus, normal decays (FS like) and late-time exponential declines\cite{Oganesyan2020,Panaitescu2020}. We have considered LAE from the jet wings, i.e. the regions outside of the jet core. The structure of the bulk Lorentz factor and the angular energy distribution follows a power-law profile with index k, i.e. $\rm \Gamma (\theta), \epsilon (\theta) \propto \theta^{-k}$ for $\rm \theta>\theta_{j}$. We have fixed the comoving spectrum to the Band function with $\rm \alpha=-1.0$ and $\rm \beta=-2.50$. Thus, the parameters that define the LAE are the following: the bulk Lorentz factor of the jet head $\Gamma_{0}$, the transparency radius $\rm R$ (the size of the jet head where jet wings are optically thin\cite{Ascenzi2020}), $E_{p}$ the "observed" peak energy of the emission, the jet opening angle $\rm \theta_{j}$ and the jet structure index k. Under our assumptions, the jet opening angle $\rm \theta_{j}$ and the power-law profile index k are common parameters for the LAE and FS models. We have included the FS and RS models together with the LAE self-consistently. The combined RS, FS and LAE model is capable of describing the complex optical and X-ray light curves. The early X-ray lightcurve is initially dominated by the LAE, while the optical light curve is entirely made by the early RS and later on FS components only. The jet of GRB~210619B is best described by the following parameters: the isotropic equivalent of the kinetic energy $\rm E_{kin} = 2.5_{-1.5}^{+10.1} \times 10^{56}$ erg, the jet opening angle of $\rm \theta_{j}=0.9_{-0.2}^{+0.1}$ deg, the power-law index of the structure $\rm k=6 \pm 2$. It corresponds to the opening angle-corrected prompt emission energy of $\rm E_{\gamma,\theta}=4.4 \pm 1.4 \times 10^{50}$ erg (consistent with the Ghirlanda relation\cite{Ghirlanda2004}), the true kinetic energy of the jet of $\rm E_{kin,\theta} = 3.8_{-2.7}^{+10.2} \times 10^{52}$ erg and the efficiency of the prompt emission of $\rm 0.013_{-0.011}^{+0.020}$. The micro-physical parameters of the FS are $\rm p=2.023_{-0.008}^{+0.009}$, $\rm \epsilon_{e,FS}=0.05_{-0.03}^{+0.05}$ and $\rm \epsilon_{B,FS}=7.9_{-7.7}^{+474} \times 10^{-4}$. The density of the circumburst medium is $\rm n = 5.5_{-4.7}^{+8.6} \times 10^{-5} \, cm^{-3}$. The RS component is peaked at $\rm 36\pm0.4$ s and declines with a temporal index of $\rm -1.75\pm0.01$ with the following exponential cutoff at $(\rm 2.3_{-0.2}^{+0.3}) \times 10^{3}$ s. The temporal decline of the RS\cite{Kobayashi2000} suggests the electron distribution index of $\rm p_{RS} \approx 2.0$. The LAE component that fits most of the X-ray emission is best represented by the following parameters: the transparency radius of the jet core $R=(8_{-1}^{+2} )\times 10^{15}$ cm, the jet core bulk Lorentz factor of $\rm \Gamma_{0}=360_{-250}^{+960}$ and the peak energy of the spectrum in the observed frame highly unconstrained and is spanning over the range of $\rm \sim 10-5600$ keV. 

{\noindent \bf The magnetisation of the jet} 
The ratio between the cooling frequencies of the RS and FS components at the shell crossing time, i.e. at $\rm t_{0,RS}$, defines roughly the relative magnetisation parameter $\rm \nu_{c,RS}/\nu_{c,FS} \approx R_{B}^{-3/2}$, where $\rm R_{B} = \epsilon_{B,RS}/\epsilon_{B,FS}$\cite{Zhang2003,Kobayashi2003}. The cutoff time suggests that the cooling frequency crosses the observed R-band at $\rm \sim 2.3 \times 10^{3}$ s. We estimate the cooling frequency of the RS component at the shell crossing time\cite{Kobayashi2000} as $\rm \nu_{R} \left(\frac{t_{0,RS}}{t_{cut,RS}} \right)^{-54/35} \sim 3 \times 10^{17}$ Hz (conservatively, in the thin shell scenario). Given the best parameters for the FS, we estimate the the cooling frequency 
at $t_{0,RS}$ as $\rm \nu_{c,FS}(t_{0,RS}) \sim 10^{18}$ Hz. Therefore, the relative magnetisation parameter is $\rm R_{B} = 2.5 \pm 0.3$ which returns the magnetic equipartition parameter of the RS $\rm \epsilon_{B,RS} = 0.12_{-0.10}^{+0.20}$. 

{\noindent \bf Markov Chain Monte Carlo sampling}
We have used the Goodman $\rm \&$ Weare’s Affine Invariant Markov chain Monte Carlo (MCMC) Ensemble sampler\cite{Goodman2010} implemented by the python package {\it emcee}\cite{Foreman-Mackey2013} to explore the parameters of our empirical and physical models. We use the sums of Gaussian log-likelihoods in order to self-consistently estimate the model parameters when considering optical and X-ray data simultaneously. We allow the MCMC to run until the number of steps exceeds fifty times of the maximum of the auto-correlation time. We report the 16th, 50th, and 84th percentiles of the samples in the marginalized distributions throughout this paper.


\newpage 

{\noindent  \bf Supplementary Materials.}

\newpage 

\newpage 

\newpage 

\begin{figure}[ht!]
    \centering
    \includegraphics[width=1.0\textwidth]{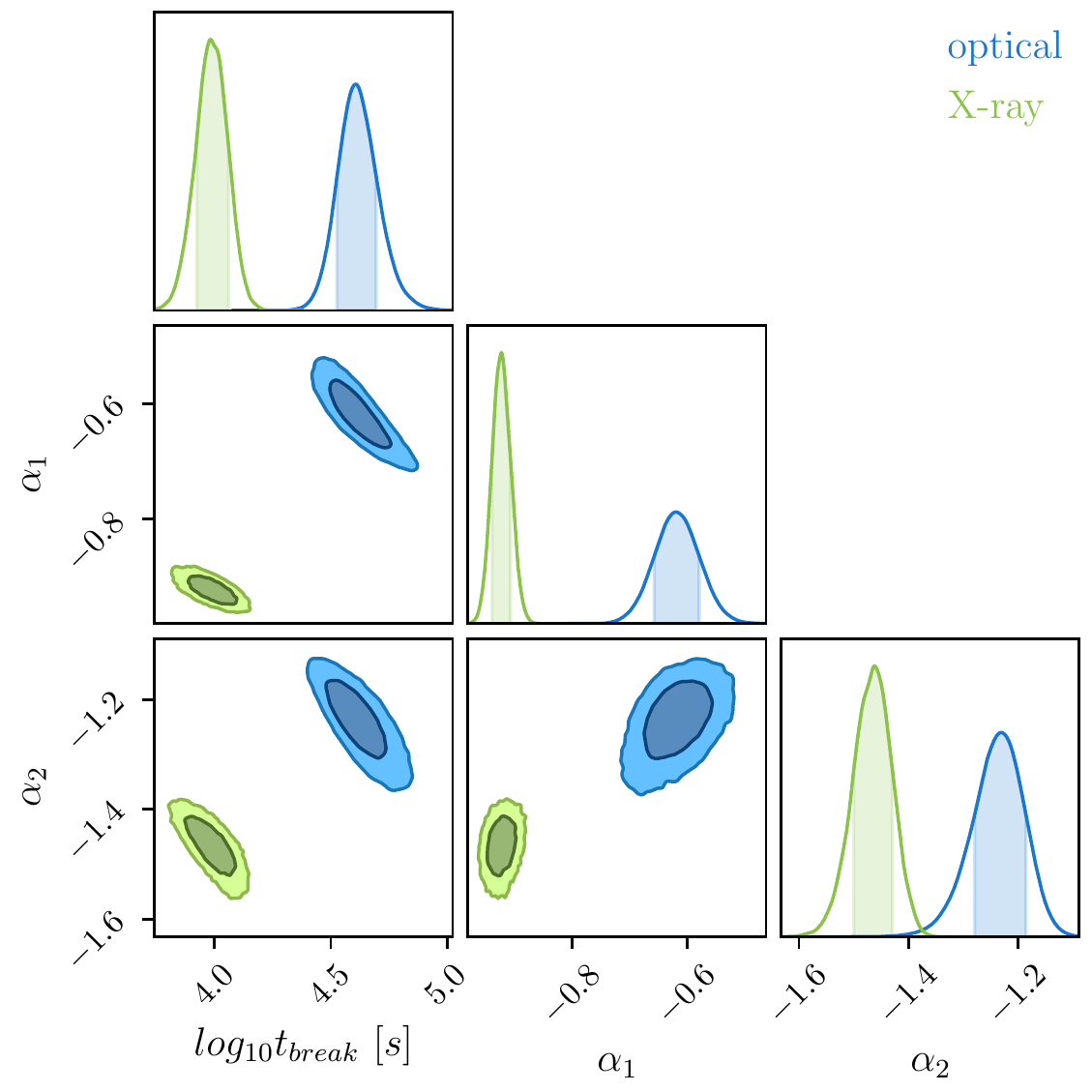}
    \caption{The joint corner plot for the parameters of the empirical model for the X-ray/optical light curves.
    }
    \label{joint_contour}
\end{figure}

\newpage

\newpage

\newpage 

\begin{figure}[ht!]
    \centering
    \includegraphics[width=1.0\textwidth]{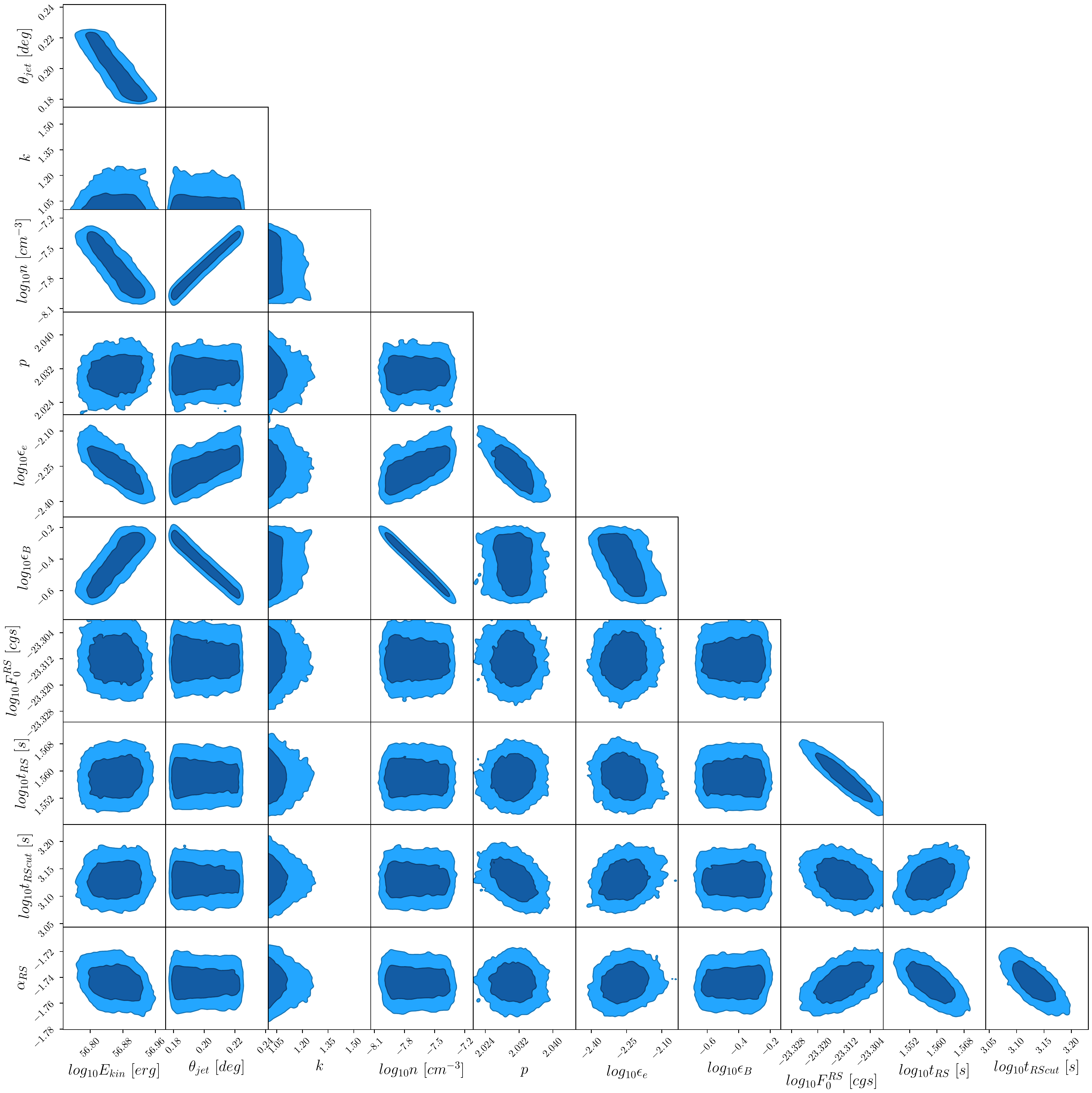}
    \caption{The corner plot for the parameters of the model which includes the emission from the reverse shock and the forward shock only.
    }
    \label{all}
\end{figure}

\newpage 

\begin{figure}[ht!]
    \centering
    \includegraphics[width=1.0\textwidth]{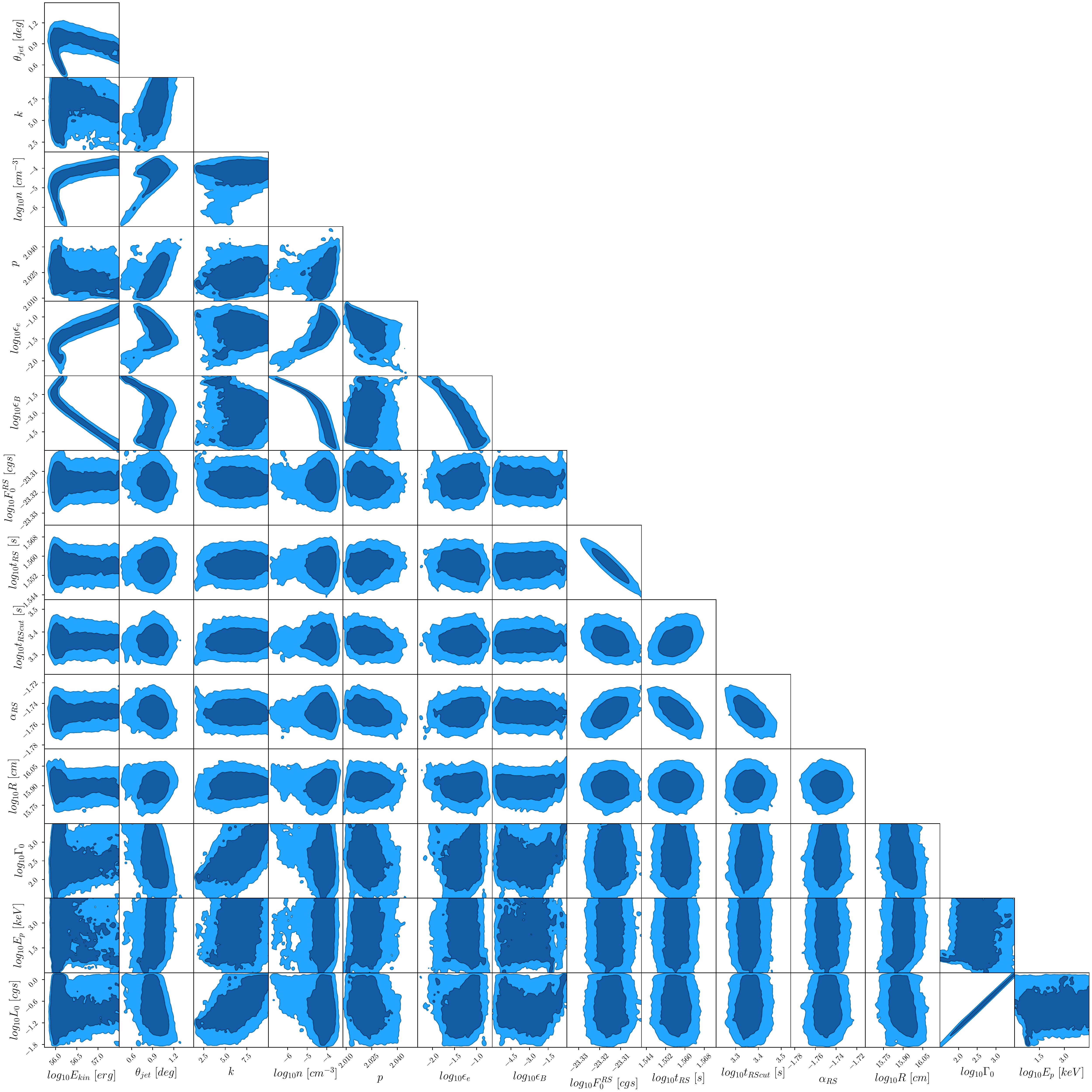}
    \caption{The corner plot for the parameters of the full model including the emission from the reverse shock, the forward shock and the Large Angle Emission.
    }
    \label{all}
\end{figure}

\newpage


\begin{table}
\centering
\begin{tabular}{lcr}
time-bin	&	observing mode 		&  Flux	\\
(s)	&	 		&    ($10^{-12}$\,erg\,cm$^{-2}$\,s$^{-1}$) 	\\
\hline	
209-283& PC & $1504.77_{-3.061}^{+2.151}$\\ 
335-500& WT & $897.72_{-0.405}^{+0.419}$\\ 
500-1000& WT & $555.66_{-0.182}^{+0.187}$\\ 
1000-1500& WT & $323.56_{-0.140}^{+0.145}$\\ 
1500-1864& WT & $283.84_{-0.162}^{+0.169}$\\ 
6395-7565& PC & $58.90_{-0.075}^{+0.083}$\\ 
11828-13237& PC & $35.42_{-0.054}^{+0.048}$\\ 
17763-19024& PC & $24.71_{-0.048}^{+0.057}$\\ 
23610-34828& PC & $8.08_{-0.010}^{+0.011}$\\ 
50936-58400& PC & $2.90_{-0.006}^{+0.007}$\\ 
58400-99305& PC & $2.20_{-0.005}^{+0.006}$\\ 
109476-196624& PC & $0.94_{-0.002}^{+0.001}$\\ 
213000-400000& PC & $0.26_{-0.001}^{+0.001}$\\ 
\hline	

\end{tabular}
\caption{Unabsorbed XRT Flux [0.5-10 keV]}
\label{xrt_tab}
\end{table}

\begin{table}
\centering
\begin{tabular}{lclllll}
time-bin	&  Optical instrument 	& $\rm \alpha$ & $\rm \beta$ & $\rm E_{0}$ & Flux \, $10^{-6}$ 	& PGSTAT/d.o.f.\\
s	&	&   & 	& keV	&   $\rm erg \, cm^{-2} \,  s^{-1}$ &	\\
\hline	
\\
55.5-56.5 & MMT-9 [channel 1] & $-0.86_{-0.22}^{+0.24}$ & $-2.66_{-1.10}^{+0.33}$ & $63.57_{-22.06}^{+26.78}$ & $2.00_{-0.33}^{+0.43}$ & 391/318 \\ 
58.7-59.7 & MMT-9 [channel 1] & $-0.38_{-0.44}^{+0.23}$ & $-2.64_{-0.71}^{+0.15}$ & $34.33_{-6.64}^{+20.92}$ & $2.50_{-0.31}^{+0.25}$ & 352/318 \\ 
60.9-61.9 & MMT-9 [channel 1] & $-0.83_{-0.17}^{+0.21}$ & $-10.00_{-0.00}^{+0.00}$ & $45.84_{-8.23}^{+9.25}$ & $1.44_{-0.06}^{+0.05}$ & 382/318 \\

\hline
\\
55.5-60.5 & MMT-9 [channel 2] & $-0.87_{-0.16}^{+0.09}$ & $-2.59_{-0.25}^{+0.10}$ & $50.07_{-12.56}^{+13.55}$ & $1.59_{-0.13}^{+0.13}$ & 392/318 \\ 
60.6-65.6 & MMT-9 [channel 2] & $-0.69_{-0.26}^{+0.24}$ & $-2.69_{-0.18}^{+0.17}$ & $31.08_{-7.52}^{+10.94}$ & $0.98_{-0.06}^{+0.09}$ & 399/318 \\   
\hline
\\
28.2-38.2 & D50 & $-0.95_{-0.07}^{+0.07}$ & $-2.55_{-0.34}^{+0.18}$ & $136.60_{-17.81}^{+22.16}$ & $2.10_{-0.21}^{+0.22}$ & 370/318 \\ 
39.3-49.3 & D50 & $-1.01_{-0.05}^{+0.05}$ & $-2.66_{-0.20}^{+0.14}$ & $119.55_{-11.09}^{+12.34}$ & $2.97_{-0.17}^{+0.18}$ & 473/318 \\ 
50.8-60.8 & D50 & $-0.91_{-0.06}^{+0.07}$ & $-2.43_{-0.10}^{+0.09}$ & $85.29_{-9.55}^{+10.04}$ & $3.08_{-0.17}^{+0.19}$ & 522/318 \\ 
62.0-72.0 & D50 & $-0.75_{-0.67}^{+0.39}$ & $-2.43_{-0.30}^{+0.13}$ & $24.55_{-7.54}^{+38.40}$ & $0.42_{-0.05}^{+0.06}$ & 388/318 \\ 

\hline	

\end{tabular}
\caption{The best fit parameters of the time-resolved \textit{Fermi}/GBM spectra modelled by the Band function. The flux is computed on the range from 8 keV to 10 MeV. }
\label{fermi_tab}
\end{table}


\begin{table}
\centering
\begin{tabular}{cccccrr}
\multicolumn{1}{c}{Interval} & FRAM Clear & MMT-9 Clear & MMT-9 V & MMT-9 B & \multicolumn{1}{c}{($g-r$)} & \multicolumn{1}{c}{$\beta$}\\
\multicolumn{1}{c}{s} & $F/F_{\rm D50}$ & $F/F_{\rm D50}$  & $F/F_{\rm D50}$  & $F/F_{\rm D50}$ &  \\
\hline
$30\ ..\ 50$ & $1.01\pm0.1$ & & & & $-0.06\pm0.75$ & $-2.10\pm2.43$\\
$50\ ..\ 100$ & $0.85\pm0.01$ & $0.71\pm0.03$ & $0.61\pm0.05$ & $<0.18$ & $0.99\pm0.07$ & $1.42\pm0.26$\\
$100\ ..\ 200$ & $0.88\pm0.01$ & $0.76\pm0.06$ & $0.62\pm0.10$ & $<0.13$ & $0.90\pm0.09$ & $1.09\pm0.33$\\
$200\ ..\ 400$ & $0.88\pm0.01$ & $0.80\pm0.12$ & $0.55\pm0.20$ & $<0.36$ & $0.88\pm0.10$ & $1.08\pm0.37$\\
$400\ ..\ 600$ & $0.90\pm0.02$ & $0.13\pm0.12$ & $0.96\pm0.44$ & $<0.65$ & $0.99\pm0.14$ & $1.53\pm0.46$\\
$600\ ..\ 1000$ & $0.83\pm0.02$ & & & & $1.34\pm0.19$ & $2.56\pm0.62$\\
\hline	
\multicolumn{7}{c}{\mbox{Late-time data}} \\
$3200$ & & & & &$0.67\pm0.07$ & $0.37\pm0.26$ \\
$5500$ & & & & &$0.57\pm0.13$ & $0.00\pm0.47$ \\
$8640$ & & & & &$0.76\pm0.04$ & $0.70\pm0.16$ \\
$13824$ & & & & &$0.82\pm0.04$ & $0.92\pm0.16$ \\
$66528$ & & & & &$0.95\pm0.09$ & $1.40\pm0.32$ \\

\hline

\end{tabular}
\caption{Average fractions of fluxes from FRAM and MMT-9 relative to D50 in different time intervals, and corresponding $(g-r)$ color indices and spectral slopes $\beta$. The fluxes and color indices are not corrected for the Galactic extinction, which contributes $E_{g-r}^G=0.17$ in the direction of the burst. The $\beta$ spectral slopes, on the other hand, are corrected for both Galactic extinction and an additional $E_{g-r}=0.40$ dust extinction due to an intervening absorber at $z=1.095$. 
For completeness, the colors and the corresponding slopes derived from the data in Tables~\ref{table:d50_filters} and \ref{table:gcns} are shown.}
\label{table:colors}
\end{table}


\begin{table}
\centering
\begin{tabular}{cccc}

Time, s & Mag & Exposure, s & Filter \\
\hline
$2771$ & $17.80\pm0.05$ & $g$ & $131$ \\
$2902$ & $17.93\pm0.06$ & $g$ & $123$ \\
$3061$ & $17.87\pm0.05$ & $g$ & $188$ \\
$3442$ & $17.20\pm0.05$ & $r$ & $60$ \\
$3505$ & $17.20\pm0.06$ & $r$ & $60$ \\
$3569$ & $17.19\pm0.06$ & $r$ & $60$ \\
$3633$ & $17.30\pm0.06$ & $r$ & $60$ \\
$3696$ & $17.30\pm0.07$ & $r$ & $60$ \\
$3760$ & $17.29\pm0.06$ & $r$ & $60$ \\
$3823$ & $17.23\pm0.06$ & $r$ & $60$ \\
$3886$ & $17.27\pm0.06$ & $r$ & $60$ \\
$3952$ & $16.88\pm0.05$ & $i$ & $60$ \\
$4014$ & $17.05\pm0.07$ & $i$ & $60$ \\
$4078$ & $16.93\pm0.06$ & $i$ & $60$ \\
$4142$ & $17.00\pm0.07$ & $i$ & $60$ \\
$4206$ & $16.91\pm0.07$ & $i$ & $60$ \\
$4271$ & $17.00\pm0.07$ & $i$ & $60$ \\
$4335$ & $17.03\pm0.07$ & $i$ & $60$ \\
$4399$ & $17.01\pm0.08$ & $i$ & $60$ \\
$4462$ & $17.02\pm0.08$ & $i$ & $60$ \\
$4589$ & $16.77\pm0.06$ & $z$ & $186$ \\
$4781$ & $16.81\pm0.08$ & $z$ & $188$ \\
$4971$ & $16.68\pm0.06$ & $z$ & $188$ \\
$5384$ & $18.14\pm0.10$ & $g$ & $632$ \\
$5799$ & $17.57\pm0.08$ & $r$ & $186$ \\
$5988$ & $17.57\pm0.09$ & $r$ & $186$ \\
$6177$ & $17.56\pm0.10$ & $r$ & $187$ \\
$6865$ & $16.90\pm0.11$ & $z$ & $1178$ \\
$84656$ & $20.16\pm0.19$ & $r$ & $180$ \\
$163842$ & $21.12\pm0.48$ & $r$ & $180$ \\

\hline	

\end{tabular}
\caption{D50 data in photometric filters}
\label{table:d50_filters}
\end{table}


\begin{table}
\centering
\begin{tabular}{cccc}

Time & Mag & Filter & Reference \\
\hline
$9498$ & $17.90\pm0.20$ & $V$ & GCN 30265\cite{gcn_30265} \\
$21492$ & $18.51\pm0.10$ & $R$ & GCN 30268\cite{gcn_30268} \\
$19824$ & $18.51\pm0.10$ & $R$ & GCN 30268\cite{gcn_30268} \\
$9490$ & $17.44\pm0.04$ & $i$ & GCN 30271\cite{gcn_30271} \\
$13848$ & $18.11\pm0.03$ & $r$ & GCN 30271\cite{gcn_30271} \\
$9307$ & $18.58\pm0.03$ & $g$ & GCN 30271\cite{gcn_30271} \\
$9399$ & $17.82\pm0.03$ & $r$ & GCN 30271\cite{gcn_30271} \\
$14149$ & $17.47\pm0.04$ & $z$ & GCN 30271\cite{gcn_30271} \\
$13939$ & $17.74\pm0.04$ & $i$ & GCN 30271\cite{gcn_30271} \\
$9700$ & $17.17\pm0.04$ & $z$ & GCN 30271\cite{gcn_30271} \\
$13755$ & $18.93\pm0.03$ & $g$ & GCN 30271\cite{gcn_30271} \\
$39852$ & $18.80\pm0.10$ & $N$ & GCN 30273\cite{gcn_30273} \\
$28980$ & $18.50\pm0.10$ & $N$ & GCN 30273\cite{gcn_30273} \\
$51948$ & $19.00\pm0.10$ & $R$ & GCN 30277\cite{gcn_30277} \\
$1318$ & $16.10\pm0.10$ & $r$ & GCN 30280\cite{gcn_30280} \\
$72180$ & $21.14\pm0.07$ & $g$ & GCN 30286\cite{gcn_30286} \\
$65700$ & $21.13\pm0.07$ & $g$ & GCN 30286\cite{gcn_30286} \\
$67284$ & $20.18\pm0.05$ & $r$ & GCN 30286\cite{gcn_30286} \\
$68904$ & $20.09\pm0.06$ & $i$ & GCN 30286\cite{gcn_30286} \\
$22104$ & $18.52\pm0.12$ & $r$ & GCN 30288\cite{gcn_30288} \\
$85679$ & $19.80\pm0.10$ & $R$ & GCN 30291\cite{gcn_30291} \\
$84154$ & $20.50\pm0.40$ & $r$ & GCN 30292\cite{gcn_30292} \\
$81215$ & $19.90\pm0.20$ & $R$ & GCN 30293\cite{gcn_30293} \\
$103699$ & $18.88\pm0.11$ & $z$ & GCN 30294\cite{gcn_30294} \\
$104881$ & $19.65\pm0.02$ & $i$ & GCN 30294\cite{gcn_30294} \\
$102244$ & $19.92\pm0.02$ & $R$ & GCN 30294\cite{gcn_30294} \\
$171876$ & $20.60\pm0.30$ & $R$ & GCN 30299\cite{gcn_30299} \\
$171289$ & $20.73\pm0.08$ & $R$ & GCN 30303\cite{gcn_30303} \\
$99792$ & $20.00\pm0.20$ & $R$ & GCN 30305\cite{gcn_30305} \\
$255989$ & $21.40\pm0.10$ & $R$ & GCN 30309\cite{gcn_30309} \\
$256068$ & $20.09\pm0.28$ & $i$ & GCN 30320\cite{gcn_30320} \\
$256068$ & $21.45\pm0.33$ & $r$ & GCN 30320\cite{gcn_30320} \\
$529067$ & $22.56\pm0.16$ & $r$ & GCN 30338\cite{gcn_30338} \\
$69088$ & $19.47\pm0.06$ & $R$ & GCN 30791\cite{gcn_30791} \\
$159856$ & $20.43\pm0.05$ & $R$ & GCN 30791\cite{gcn_30791} \\
$236914$ & $21.82\pm0.26$ & $r$ & GCN 30791\cite{gcn_30791} \\
$241580$ & $21.21\pm0.09$ & $R$ & GCN 30791\cite{gcn_30791} \\
$593153$ & $22.55\pm0.24$ & $R$ & GCN 30791\cite{gcn_30791} \\
$1111648$ & $23.50\pm0.30$ & $R$ & GCN 30791\cite{gcn_30791} \\
\hline	

\end{tabular}
\caption{Photometric data collected from GCN Circulars}
\label{table:gcns}
\end{table}

\newpage

\end{document}